\documentclass{aa}

\def\vec#1{\ensuremath{\mathchoice{\mbox{\boldmath$\displaystyle#1$}}
{\mbox{\boldmath$\textstyle#1$}}
{\mbox{\boldmath$\scriptstyle#1$}}
{\mbox{\boldmath$\scriptscriptstyle#1$}}}}

\usepackage[varg]{txfonts}

\newcommand{\glon}{\mathfrak{L}}
\newcommand{\glat}{\mathfrak{B}}

\begin{document}

\title{Rigorous treatment of barycentric stellar motion}
\subtitle{Perspective and light-time effects in astrometric and radial velocity data}

\titlerunning{Rigorous treatment of barycentric stellar motion}

\author{A. G. Butkevich \inst{1,2} \and L. Lindegren\inst{3}}

\institute{Lohrmann Observatory, Technische Universit\"at Dresden,
                01062 Dresden, Germany
           \and Pulkovo Observatory, Pulkovskoye shosse 65, 196140
                Saint-Petersburg, Russia\\
           \email{alexey.butkevich@tu-dresden.de}
         \and
           Lund Observatory, Box 43, 22100 Lund, Sweden\\
           \email{lennart@astro.lu.se}}

\date{Received; accepted}

\abstract
{High-precision astrometric and radial-velocity observations require accurate modelling of
stellar motions in order to extrapolate measurements over long time intervals, and to detect
deviations from uniform motion caused, for example, by unseen companions.}
{We aim to explore the simplest possible kinematic model of stellar motions, namely that of
uniform rectilinear motion relative to the solar system barycentre, in terms of observable
quantities including error propagation.}
{The apparent path equation for uniform rectilinear motion is solved analytically in a classical
(special-relativistic) framework, leading to rigorous expressions that relate the (apparent)
astrometric parameters and radial velocity to the (true) kinematic parameters of the star in
the barycentric reference system.}
{We present rigorous and explicit formulae for the transformation of stellar positions, parallaxes,
proper motions, and radial velocities from one epoch to another, assuming uniform rectilinear
motion and taking light-time effects into account. The Jacobian matrix of the transformation
is also given, allowing accurate and reversible propagation of errors over arbitrary time intervals.
The light-time effects are generally very small, but exceed 0.1~mas or 0.1~m~s$^{-1}$ over
100~yr for at least 33 stars in the Hipparcos catalogue. For high-velocity stars within a few tens
of pc from the Sun, light-time effects are generally more important than the effects of the
curvature of their orbits in the Galactic potential.}
{}


\maketitle

\section{Introduction}

The pioneering Hipparcos mission necessitated many refinements in the
analysis of astrometric observations. Effects that were previously ignored when constructing
stellar catalogues, such as gravitational light deflection by bodies in the solar system and
relativistic stellar aberration, had to be systematically taken into account in order to reach
the milli-arcsecond (mas) accuracy made possible by observations from space. The Gaia mission,
aiming at positional accuracies at the 10~micro-arcsecond ($\mu$as) level \citep{deBruijne2012},
requires further sophistication of data modelling to account for the subtle physical effects that
come into play at this accuracy. A practical model for the relativistic reduction of astrometric
observations, accurate to 1~$\mu$as, was formulated by \citet{klioner2003} and is the basis
for the astrometric processing of the Gaia data \citep{lindegrenAGIS2012}.

A basic assumption in these models is that stars move with constant velocity (speed and
direction) relative to the solar system barycentre (SSB). For binaries and other non-single systems,
including exoplanetary systems, their centres of mass are instead assumed to move with
uniform velocity. The assumption, referred to here as the uniform rectilinear model,
is fundamental in several respects. First of all, it allows us to describe the motion
of any star, or the centre of mass of a multi-body system, compactly in terms of a handful of
easily catalogued parameters. Secondly, it allows us to extrapolate their motions forwards and backwards
in time in order to serve as comparison points for observations at arbitrary epochs. Thirdly, it provides
a reference model, or null hypothesis, for the detection of non-linear motions caused for example
by planetary companions. Indeed, the uniform rectilinear model is used as a reference in all modern
observational analysis of stellar motions, including non-astrometric techniques such as
high-precision Doppler monitoring \citep[e.g.][]{choi+2013}. In the analysis of pulsar timings,
the curvature of Galactic stellar orbits is taken into account \citep{edwards+2006}, but then
only as a known correction to the uniform motion.

In this paper, we consider the application of the uniform rectilinear model to the problem of propagating
the astrometric parameters from one epoch to another, which is a very common task in the practical
use of such data. Various aspects of this problem have been extensively discussed by several
authors (see Sect.~\ref{sec:prev}), but we provide for the first time a rigorous analytical solution
including the propagation of uncertainties, as given by the covariance matrix of the astrometric
parameters. Physical limitations of the uniform rectilinear model are discussed in
Appendix~\ref{ss:model_applicability}.

It has long been recognized that the accurate propagation of stellar positions
needs to take radial motions as well as the tangential (proper) motions into account.
Thus radial velocity is inextricably connected with astrometric data and is sometimes
regarded as the ``sixth astrometric parameter'', complementing the standard five (for position,
parallax, and proper motion) in defining stellar coordinates in six-dimensional phase space.
In the present paper we adopt this view even though the radial motion is usually determined
by the spectroscopic method.

The astrometric parameters of a star are derived from observations using established
formulae, as detailed by \cite{klioner2003}, to correct for local effects such as gravitational
deflection and the position and motion of the observer. Similarly,
spectroscopic Doppler measurements need to be corrected for local and astrophysical
effects as described by \citet{lindegren_dravins2003}. Effectively, the result is a set of
parameters describing the observed phenomena as they would appear for a fictitious
observer located at the SSB in the absence of the gravitational fields of all solar system
bodies. The subsequent analysis of the observations can then entirely be made in a
classical (or special relativistic) framework.

The exact relation between the uniform rectilinear model and the astrometric parameters
(including radial velocity) is simple in principle but rather more complicated in practice,
primarily owing to the vastly different uncertainties in the radial and
tangential components of stellar coordinates. While stellar distances are seldom
known to a relative precision better than $10^{-2}$, their angular coordinates may be
determined at least six orders of magnitude more accurately. This has two important
consequences. First, that astrometric observations cannot easily be modelled directly
in terms of the rectangular coordinates of stellar positions and velocities. The
astrometric parameters, using spherical coordinates and parallax, were introduced to
overcome this difficulty. Secondly,
because the light-travel time from the star to the observer is generally not very well
known, it is customary, and in practice necessary, to define the astrometric parameters
as apparent quantities by effectively ignoring the light-time effects. The resulting
relation between the physical model and observed quantities therefore includes
both the classical geometric effects and those due to the finite speed of light.

After a brief historical review we discuss the general effects of the light-travel time in
Section~\ref{s:light-time}. Section~\ref{s:astr_params} presents some prerequisite material.
Section~\ref{s:including} contains an analytical treatment of the epoch transformation. Results and conclusions are summarized in Sects.~\ref{s:discussion}
and \ref{s:conclusion}.

\section{Previous work}\label{sec:prev}

As described by \citet{schlesinger1917}, a star's movement through space causes secular
changes not only in the position, but also in its proper motion, parallax, and radial velocity
as observed from the Sun (or the SSB). These are purely geometric effects due to the
changing distance and angle between the line of sight and the direction of motion.
Schlesinger proposed that the resulting quadratic term in angular position, known as the
secular (or perspective) acceleration, could be used to determine the radial velocities
of some stars ``independently of the spectroscope and with an excellent degree of precision.''
This has so far only been possible for very few stars \citep{vanDeKamp1977,dravins+lindegren1999}.
Nevertheless, the work pioneered the use of the uniform rectilinear model for propagating
astrometric data over longer intervals of time.
Traditionally, the propagated quantities are represented by series expansions in time, leading
to well-known formulae for the secular acceleration and time derivative of proper motion
\citep{scott_hughes1964, mueller1969, taff1981, murray1983}. The drawback of this approach
is that its applicability is limited to a certain timespan, depending on the required accuracy
and the sizes of neglected terms. This can be avoided by using transformations that directly
link the spherical coordinates at the different epochs by means of Cartesian vectors. This also
leads to a considerable simplification of the mathematical formulation of the problem. To our
knowledge, this approach was pioneered by \cite{eichhorn_rust1970}, who derived
expressions for the variations in proper motion valid for any, not necessarily small, time interval.
The procedure yields a straightforward propagation formula,
which came into common use in the Hipparcos data reduction \citep{lindegren_et_al1992} and
formed a theoretical basis for the semi-rigorous treatment of the epoch transformation
developed by \citet{lindegren1995b} and subsequently published in the Hipparcos catalogue
\citep[][Vol.~1, Sect.~1.5.5]{esa1997}. With respect to the uniform rectilinear model it is
semi-rigorous in the sense that light-time effects are ignored, as they could be
shown to be negligible at the Hipparcos level of accuracy.

Possibly the first treatment of astrometric light-time effects was by \citet{schwarzschild1894},
who in a discussion of ``secular aberration'' (stellar aberration due to the motion of the solar
system) derived a relation between the apparent and true proper motions correct to first
order in $v/c$. In a largely overlooked paper by \citet{eisner1967} the rigorous propagation
was derived as a series expansion in $t$ and $v/c$, though in a form not very useful for
practical application. The author concluded that the light-time
effect ``can be neglected in astrometry of present-day precision, though not necessarily if
measurement outside the atmosphere becomes practical.'' The most complete analysis of the
problem so far was carried
out by \citet{stumpff1985}, who derived the rigorous relations between the apparent and true
quantities, based on the uniform rectilinear model including light-time effects.
The present work extends the treatment by \citet{stumpff1985} in several respects, as
discussed in Sect.~\ref{ss:stumpff}.

Since knowledge of uncertainties is essential for the exploitation of astrometric data, the
epoch transformation must be accompanied by the associated error propagation.
Strictly speaking, because of correlation between data items, uncertainties as such are not
meaningful in this context and covariances should be used instead. The procedure for transforming
astrometric data developed consistently by \cite{lindegren1995b} includes the propagation of the
associated covariance matrix, but without light-time effects. In the present work,
we generalize this technique by incorporating the effects of the finite light-travel time.

\section{Light-time effects for the uniform motion}\label{s:light-time}

\subsection{The uniform rectilinear model}

According to the uniform rectilinear model, the barycentric vector of the star at
the arbitrary epoch $T$ is given by
\begin{equation}\label{eq:classical-equation-motion}
 \vec{b}(T)=\vec{b}_0+(T-T_0)\,\vec{v}\,,
\end{equation}
where $\vec{b}_0$ is the barycentric position at the initial epoch $T_0$ and
$\vec{v}$ the constant space velocity. The model has six kinematic parameters,
namely the components of vectors $\vec{b}_0$ and $\vec{v}$ in the barycentric
reference frame. The conditions of applicability of this model are considered in
Appendix~\ref{ss:model_applicability}.

An equivalent form of Eq.~(\ref{eq:classical-equation-motion}) is obtained by
considering two distinct moments of time, distinguished by subscript 1 and 2:
\begin{equation}\label{eq:classical-equation-motion-diff}
 \vec{b}(T_2)=\vec{b}(T_1)+(T_2-T_1)\,\vec{v} \, .
\end{equation}

As emphasized in the introduction, we work here in a special-relativistic framework
since all the effects of general relativity can be assumed to have been taken into
account in the reduction from measurable (proper) directions to coordinate directions, as comprehensively discussed by
\citet{klioner2003}. From here on, by observation we mean the information about the
instantaneous position and velocity of a star as seen by an observer at the SSB,
referring to a specific moment read by observer's clock. Thus, we ignore all
practical aspects of observation and reduction of astrometric data. Indeed,
since the observer is assumed to be at rest at the origin (SSB) we do not even
need to consider the special-relativistic transformation between different
observers: all derivations can be made in an entirely classical way using the
constant coordinate speed of light to take into account light-time effects.

\begin{figure}[t]
\resizebox{\hsize}{!}{\includegraphics{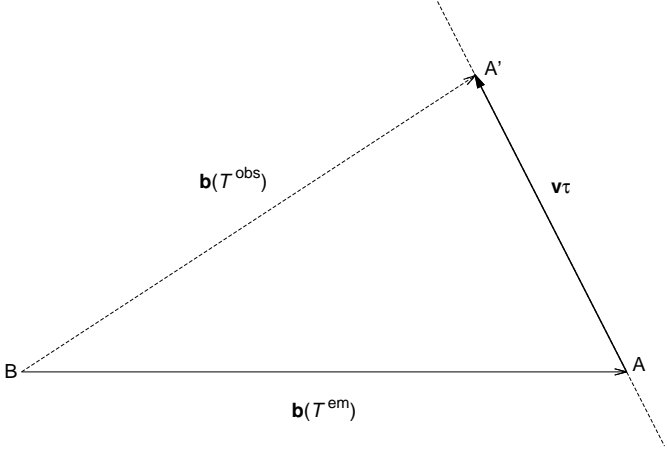}}
% \resizebox{\hsize}{!}{\includegraphics{fig1.pdf}}
  \caption{Light-time effects for the observation of a uniformly moving star
by an observer at the solar system barycentre B. The plot explicitly demonstrates
the distinction between the apparent and true position of the star described
by Eq.~(\ref{eq:pos-relation}). The apparent position A, observed at time
$T^\mathrm{obs}$, is given by the vector $\vec{b}(T^\mathrm{em})$. During the
time it takes for the light to travel from A to B the star has moved from A to A'.}
  \label{fig:1}
\end{figure}

\subsection{The light-time equation}

The finite speed of light makes it necessary to distinguish between the time
$T^\mathrm{em}$ when a light signal was emitted by a star, and the time
$T^\mathrm{obs}$ when the same signal was detected by the observer.
The two moments in time are connected by the light-time equation
\begin{equation}\label{eq:light-time}
    T^\mathrm{obs} = T^\mathrm{em}+b(T^\mathrm{em})/c \, ,
\end{equation}
where $b$ is the barycentric distance of the star.

Using the uniform rectilinear model in Eq.~(\ref{eq:classical-equation-motion})
and taking the difference between the barycentric vectors at the two moments
of time yields
\begin{equation}\label{eq:pos-relation}
    \vec{b}(T^\mathrm{em})=\vec{b}(T^\mathrm{obs})-\tau\vec{v} \, ,
\end{equation}
where we have introduced the light-travel time, or light-time,
\begin{equation}\label{eq:light-travel-time-definition}
    \tau=T^\mathrm{obs}-T^\mathrm{em} \, .
\end{equation}
Equation~(\ref{eq:pos-relation}), illustrated in Fig.~\ref{fig:1}, corresponds
to the well-known ``planetary aberration'' effect in classical astronomy
\citep[e.g.][]{woolard+clemence1966}.

\subsection{True and apparent quantities}

The important point of Eq.~(\ref{eq:pos-relation}) is that the direction to the
star at the time of observation, $T^\mathrm{obs}$, is given by the barycentric
vector $\vec{b}(T^\mathrm{em})$ at the earlier time $T^\mathrm{em}$.
The position at the time of observation, $\vec{b}(T^\mathrm{obs})$, is not
directly observable (at least not at time $T^\mathrm{obs}$, although it might
be inferred by means of Eq.~\ref{eq:pos-relation}).

This fact suggests that we need to recognize the difference between observable
quantities, such as $\vec{b}(T^\mathrm{em})$ at time $T^\mathrm{obs}$, and
those that cannot be directly observed, such as $\vec{b}(T^\mathrm{obs})$.
We shall refer to the observable quantities as apparent, while the
unobservable quantities are referred to as true. Thus we may write
\begin{equation}\label{eq:true-apparent}
   \vec{b}^\mathrm{app}(T^\mathrm{obs}) \equiv \vec{b}^\mathrm{true}(T^\mathrm{em})
   \quad\quad \left[\,= \vec{b}(T^\mathrm{em})\,\right] \, .
\end{equation}
The bracketed equality emphasizes that the quantities considered up until now have
all been true in the above sense.

The uniform rectilinear model in Eq.~(\ref{eq:classical-equation-motion})
or (\ref{eq:classical-equation-motion-diff}) is of course expressed entirely in terms
of true quantities. We shall now re-write it in terms of apparent quantities at the
times of observations. Since $\tau=b^\mathrm{app}(T^\mathrm{obs})/c$ we
consider the light-time $\tau$ to be an observable (apparent) quantity.

If $T_1$ and $T_2$ in Eq.~(\ref{eq:classical-equation-motion-diff}) are taken to be
the times of emission we can write
\begin{equation}\label{eq:classical-equation-motion-diff1}
 \vec{b}^\mathrm{true}(T_2^\mathrm{em})=\vec{b}^\mathrm{true}(T_1^\mathrm{em})
 +(T_2^\mathrm{em}-T_1^\mathrm{em})\,\vec{v}^\mathrm{true} \, .
\end{equation}
Using Eq.~(\ref{eq:true-apparent}) to re-write the left-hand side, and
Eq.~(\ref{eq:light-travel-time-definition}) to express
$T_2^\mathrm{em}-T_1^\mathrm{em}$ in terms of observable quantities, we find
\begin{equation}\label{eq:pos-t-em-t-obs}
    \vec{b}^\mathrm{app}(T_2^\mathrm{obs})
    =
    \vec{b}^\mathrm{app}(T_1^\mathrm{obs})
    +
    \left[
    (T_2^\mathrm{obs}-T_1^\mathrm{obs}) - (\tau_2-\tau_1)
    \right] \vec{v}^\mathrm{true} \, .
\end{equation}
This almost achieves our goal, except that the formula still contains one quantity,
$\vec{v}^\mathrm{true}$, that cannot be directly observed.

The (true) velocity is by definition the time derivative of the position vector, as is also
evident from Eq.~(\ref{eq:classical-equation-motion}). It does not matter whether we
use the time of emission or observation when calculating the derivative, as long as the
same time is used as argument of the position vector being differentiated; that is
\begin{equation}\label{eq:vel-true}
    \vec{v}^\mathrm{true}
    =
    \frac{\mathrm{d}\vec{b}^\mathrm{true}(T^\mathrm{em})}
         {\mathrm{d}T^\mathrm{em}}
    =
    \frac{\mathrm{d}\vec{b}^\mathrm{true}(T^\mathrm{obs})}
         {\mathrm{d}T^\mathrm{obs}} \, .
\end{equation}
Actually, none of these derivatives can be directly measured. The only velocity that can,
in principle, be obtained directly from observations is the derivative of the apparent position
with respect to the time of observation, or
\begin{equation}\label{eq:vel-app}
    \vec{v}^\mathrm{app}
    =
    \frac{\mathrm{d}\vec{b}^\mathrm{app}(T^\mathrm{obs})}
         {\mathrm{d}T^\mathrm{obs}}
    =
    \frac{\mathrm{d}\vec{b}^\mathrm{true}(T^\mathrm{em})}
         {\mathrm{d}T^\mathrm{obs}}
    \,.
\end{equation}
Comparison with Eq.~(\ref{eq:vel-true}) shows that the velocities are related through
\begin{equation}\label{eq:vel-app-true-start}
    \vec{v}^\mathrm{app}
    =
    \vec{v}^\mathrm{true}
    \frac{\mathrm{d}T^\mathrm{em}}
         {\mathrm{d}T^\mathrm{obs}}
\end{equation}
and
\begin{equation}\label{eq:vel-true-app-start}
    \vec{v}^\mathrm{true}
    =
    \vec{v}^\mathrm{app}
    \frac{\mathrm{d}T^\mathrm{obs}}
         {\mathrm{d}T^\mathrm{em}} \, .
\end{equation}
To proceed, we need expressions for these derivatives such that
$\mathrm{d}T^\mathrm{em}/\mathrm{d}T^\mathrm{obs}$ is written as a function of
the true velocity, while $\mathrm{d}T^\mathrm{obs}/\mathrm{d}T^\mathrm{em}$ is
written as a function of the apparent velocity.

These expressions are obtained from the light-time equation (\ref{eq:light-time}).
Differentiating with respect to $T^\mathrm{em}$ gives
\begin{equation}\label{eq:dtobs-dtem}
    \frac{\mathrm{d}T^\mathrm{obs}}
         {\mathrm{d}T^\mathrm{em}}
    =
    1+v_r^\mathrm{true}/c \, ,
\end{equation}
where we have defined the true radial velocity%
\footnote{Called ``kinematic radial velocity'' in \citet{lindegren_dravins2003}.}
as the derivative of the (true) barycentric distance with respect to the time of light emission,
\begin{equation}\label{eq:vr-true}
    v_r^\mathrm{true} =
    \biggl. \frac{\mathrm{d}b^\mathrm{true}(T)}{\mathrm{d}T} \, \biggr|_{\, T=T^\mathrm{em}} \, .
\end{equation}
By writing $\vec{b}=\vec{u}b$, where $\vec{u}$ is a unit vector, we have
$\mathrm{d}b/\mathrm{d}T=\vec{u}'(\mathrm{d}\vec{b}/\mathrm{d}T)$, from which it
follows that $v_r^\mathrm{true}$ is the projection of $\vec{v}^\mathrm{true}$ along the
line-of-sight $\vec{u}^\mathrm{true}(T^\mathrm{em})=\vec{u}^\mathrm{app}(T^\mathrm{obs})$.%
\footnote{The expression on the right-hand side of Eq.~(\ref{eq:dtobs-dtem}) is
sometimes referred to as the ``Doppler factor'' \citep[e.g.][]{stumpff1985}. The name
derives from the circumstance that in the classical approximation it equals the ratio
of observed to rest-frame wavelengths, $\lambda_\mathrm{obs}/\lambda_\mathrm{lab}$,
for a non-moving observer. In special relativity the wavelength ratio obtains an additional
(Lorentz) factor due to the different rates of $T^\mathrm{em}$ and the proper time at
the star; this factor depends on the total velocity $\vec{v}^\mathrm{true}$, and therefore
involves its tangential component as well as the radial.}

The expression for $\mathrm{d}T^\mathrm{obs}/\mathrm{d}T^\mathrm{em}$ is obtained
in a similar way. Differentiating Eq.~(\ref{eq:light-time}) with respect to $T^\mathrm{obs}$,
while using Eq.~(\ref{eq:true-apparent}), gives
\begin{equation}\label{eq:dtem-dtobs}
    \frac{\mathrm{d}T^\mathrm{em}}
         {\mathrm{d}T^\mathrm{obs}}
    =
    1-v_r^\mathrm{app}/c \, ,
\end{equation}
where we have defined the apparent radial velocity%
\footnote{Called ``astrometric radial velocity'' in \citet{lindegren_dravins2003}.}
as the derivative of the (apparent) barycentric distance with respect to the time of observation,
\begin{equation}\label{eq:vr-app}
    v_r^\mathrm{app} =
    \biggl. \frac{\mathrm{d}b^\mathrm{app}(T)}{\mathrm{d}T} \, \biggr|_{\, T=T^\mathrm{obs}} \, .
\end{equation}
It is readily verified that $v_r^\mathrm{app}$ is the projection of $\vec{v}^\mathrm{app}$
along the line-of-sight $\vec{u}^\mathrm{app}$ at the time of observation.

Substituting Eqs.~(\ref{eq:dtobs-dtem}) and (\ref{eq:dtem-dtobs}) into Eqs.~(\ref{eq:vel-app-true-start}) and
(\ref{eq:vel-true-app-start}), respectively, we obtain the transformations between the apparent and true velocities
\begin{align}
 \label{eq:vel-app-true}
    \vec{v}^\mathrm{app}
    &=
    \vec{v}^\mathrm{true} \left( 1+v_r^\mathrm{true}/c \right)^{-1} \, , \\
 \label{eq:vel-true-app}
    \vec{v}^\mathrm{true}
    &=
    \vec{v}^\mathrm{app} \left( 1-v_r^\mathrm{app}/c \right)^{-1} \, .
\end{align}
Thus, the velocities are related through the Doppler factor (or its inverse);
the directions of the apparent and true velocities are the same, while their
absolute values are different.

While the true velocity is constant, according to the uniform rectilinear model,
its radial component $v_r^\mathrm{true}=\vec{u}(T)'\vec{v}^\mathrm{true}$
in general changes gradually as the star moves along a straight line because of the
changing line-of-sight direction $\vec{u}(T)$. As shown by Eq.~(\ref{eq:vel-app-true})
this means that the apparent velocity, in general, is also a function of time.
The exception is for a star without proper motion, i.e. moving along a straight
line passing through the observer, in which case the true radial velocity is constant.

We are now in position to write down the kinematic model entirely in terms of
apparent quantities. Inserting Eq.~(\ref{eq:vel-true-app}) in (\ref{eq:pos-t-em-t-obs})
we obtain:
\begin{multline}\label{eq:apparent-path}
    \vec{b}^\mathrm{app}(T_2^\mathrm{obs})
    =
    \vec{b}^\mathrm{app}(T_1^\mathrm{obs})
    + \\
    \left[ (T_2^\mathrm{obs}-T_1^\mathrm{obs}) - (\tau_2-\tau_1) \right]
    \frac{\vec{v}^\mathrm{app}}{1-v_r^\mathrm{app}/c} \, .
\end{multline}
As already mentioned, the apparent velocities $\vec{v}^\mathrm{app}$ and
$v_r^\mathrm{app}$, which appear in the final factor of Eq.~(\ref{eq:apparent-path}),
are in general functions of time. However, since this factor equals the true velocity,
which is independent of time, it can be evaluated for any time including
$T_1^\mathrm{obs}$ and $T_2^\mathrm{obs}$.

Equation~(\ref{eq:apparent-path}), governing the apparent path of the star, is
fundamental for calculating the apparent quantities at an arbitrary moment of time.
We discuss its solution in Sect.~\ref{s:including}. The classical path equation
(\ref{eq:classical-equation-motion-diff}) is recovered in the limit as $c\to\infty$.

\section{Astrometric parameters}\label{s:astr_params}

In this section, we define the astrometric parameters complying with the
kinematic model described above and introduce the corresponding
notations.

The instantaneous kinematic state of the star in the
barycentric frame is conventionally specified by means of six parameters. All
six parameters can, at least in principle, be derived from observations made
from a platform in orbit around the solar system barycentre, such as the Earth
or a satellite. The parameters are therefore observable or apparent in the sense
discussed above, and they refer to the time of observation $T^\mathrm{obs}$.

From here on, nearly all quantities discussed in this paper are in fact apparent,
and the time used is that of the observation. For brevity, we can therefore
omit the subscripts ``true'', ``app'', ``em'', and ``obs'' in most equations,
and only use them where they are needed to avoid ambiguity. Their absence
thus implies an apparent or observed quantity.

Five of the six astrometric parameters are the classical parameters:
right ascension $\alpha$, declination $\delta$, trigonometric parallax $\varpi$,
proper motion in right ascension $\mu_{\alpha*}$, and proper motion in declination
$\mu_\delta$. The sixth parameter could be the ``astrometric radial velocity'' $v_r$
\citep{lindegren_dravins2003}, equivalent to the ``apparent radial velocity'' of \citet{klioner2003},
but for reasons that will become clear later we prefer to use the ``radial proper motion''
\begin{equation}\label{eq:mur}
\mu_r = v_r\varpi/A
\end{equation}
\citep{lindegrenAGIS2012}. Here $A$ is the astronomical unit \citep{iau2012B2}.
All six parameters are barycentric in the sense that they are derived from observations,
which by necessity are non-barycentric, through the application of various corrections, so that
they effectively refer to a fictitious observer at the barycentre.
Similarly $T$ is the (fictitious) time
of light reception at the barycentre. For the precise definition of the parameters in a
general-relativistic framework and an exposition of the relevant corrections, we refer to
\citet{klioner2003}. For stellar observations, the end result of this process is a set of astrometric
parameters that, to sufficient accuracy, can be interpreted in a completely classical way, as we
do in this paper (cf.\ Appendix~\ref{ss:model_applicability}). The timescale for $T$ is barycentric coordinate time, TCB.

The six astrometric parameters $\alpha$, $\delta$, $\varpi$, $\mu_{\alpha*}$, $\mu_\delta$,
$\mu_r$ change continuously with $T$ due to the space motion of the star.
Therefore, a ``reference epoch'' $T_0$ must be chosen, purely as a matter of convention, and
we shall subsequently use $t=T-T_0$ as the time argument in all expressions instead of $T$.
Furthermore, to simplify the expressions we often omit the time argument but use subscript 0
to denote quantities at $t=0$ and the corresponding unsubscripted variables when they refer
to an arbitrary $t$.

We shall now give a precise definition of the six astrometric parameters in terms of the quantities
introduced in Sect.~\ref{s:light-time}. The barycentric vector $\vec{b}$ is not directly
observable but barycentric coordinate direction, given by the unit vector
\begin{equation}\label{eq:u}
\vec{u} = \frac{\vec{b}}{\left| \vec{b} \right|} \, ,
\end{equation}
is observable, and so is its time derivative, the proper motion vector
\begin{equation}\label{eq:mu}
\vec{\mu} = \frac{\mathrm{d}\vec{u}}{\mathrm{d}t} \, .
\end{equation}
Both are normally expressed in the ICRS. Although these vectors together have six coordinate
components, they must at any time satisfy two scalar constraints, namely $\vec{u}'\vec{u}=1$
and $\vec{u}'\vec{\mu}=0$, and so have four degrees of freedom. They correspond to the four
astrometric parameters $\alpha$, $\delta$, $\mu_{\alpha*}$, and $\mu_\delta$. To obtain
the first two parameters, let $\vec{r}$ be a fixed unit vector coinciding with $\vec{u}$
at the given time $t$. Its coordinates in ICRS are
\begin{equation}\label{eq:r}
\vec{r}=
\begin{pmatrix}
\cos\delta\cos\alpha \\
\cos\delta\sin\alpha \\
\sin\delta
\end{pmatrix}\, ,
\end{equation}
from which $\alpha$ and $\delta$ are obtained.%
\footnote{The distinction between $\vec{r}$ and $\vec{u}$ may at first seem pointless or at
least over-pedantic. In fact, as explained in Sect.~\ref{sec:propCov}, it is relevant for the
interpretation of the proper motion components and their uncertainties.\label{fn:ru}}
However, while $\vec{r}$ is uniquely given by $\alpha$ and $\delta$, the reverse is not true:
any given $\vec{r}$ can be represented by an infinite set of $(\alpha,\,\delta)$ pairs.
Restricting their ranges (e.g. $0\le\alpha<2\pi$ and
$-\pi/2\le\delta\le\pi/2$) removes most
of the ambiguity, but in the special cases of $\delta=\pm\pi/2$
(exactly) the choice of
$\alpha$ is still arbitrary. Nevertheless, a choice must be made, as it determines the
subsequent calculation of the proper motion components $\mu_{\alpha*}$ and $\mu_\delta$
from $\vec{\mu}$.
Given $\alpha$ and $\delta$ we can calculate the unit vectors in the directions of
increasing right ascension and declination, which are
\begin{equation}\label{eq:pq}
\vec{p}=
\begin{pmatrix}
-\sin\alpha \\
\cos\alpha \\
0
\end{pmatrix}\, ,\quad
\vec{q}=
\begin{pmatrix}
-\sin\delta\cos\alpha \\
-\sin\delta\sin\alpha \\
\cos\delta
\end{pmatrix}\,,
\end{equation}
respectively. Equations~(\ref{eq:r})--(\ref{eq:pq}) define three orthogonal unit vectors forming the so-called
normal triad $[\vec{p}~\vec{q}~\vec{r}]$ at $\vec{r}$ relative to the celestial reference frame
\citep{murray1983}. We now define the proper motion components as the coordinates of $\vec{\mu}$
along the $\vec{p}$ and $\vec{q}$ axes, or
\begin{equation}\label{eq:proper_motion}
\mu_{\alpha*}=\vec{p}'\vec{\mu}\, , \quad \mu_\delta=\vec{q}'\vec{\mu} \, .
\end{equation}
Conversely, since $\vec{r}'\vec{\mu}=0$, the proper motion vector can be reconstructed as
\begin{equation}\label{eq:mu_from_comp}
\vec{\mu} = \vec{p}\mu_{\alpha*} + \vec{q}\mu_\delta \, .
\end{equation}
We note that the procedure above allows us to calculate the proper motion components even when
$\delta=\pm\pi/2$, using the arbitrarily chosen value of $\alpha$: the latter defines the directions
of the $\vec{p}$ and $\vec{q}$ vectors according to Eq.~(\ref{eq:pq}) and therefore the resulting
values of $\mu_{\alpha*}$ and $\mu_\delta$ from Eq.~(\ref{eq:proper_motion}).

An alternative interpretation of the proper motion components is
\begin{equation}\label{eq:proper_motion_alt}
\mu_{\alpha*}=\frac{\mathrm{d}\alpha}{\mathrm{d}t}\cos\delta \, , \quad
\mu_\delta=\frac{\mathrm{d}\delta}{\mathrm{d}t} \, .
\end{equation}
It is readily verified that this is equivalent to
Eq.~(\ref{eq:proper_motion}) when
$\left|\delta\right|<\pi/2$, but at the celestial poles it obviously fails. We therefore
regard Eq.~(\ref{eq:proper_motion}) as the more general interpretation.

Stellar parallax is sometimes defined as the angle subtended by 1~au at the star's distance
from the Sun \citep[e.g.][]{2001eaa}. Interpreting ``distance'' to mean the coordinate
distance $b=\left|\vec{b}\right|$ from the solar system barycentre, this definition
is still ambiguous as to the precise relation to parallax $\varpi$: it could be
$\sin\varpi=A/b$ \citep{murray1983}, $\tan\varpi=A/b$ \citep{bm1998}, or
even $2\sin(\varpi/2)=A/b$ (if the astronomical unit is the chord of the angle);
the differences, of the order of $\varpi^3<10^{-10}$~arcsec,
are truly negligible for all stars. Following \citet{klioner2003} we adopt the
mathematically simplest relation
\begin{equation}\label{eq:varpi}
\varpi = \frac{A}{b} \, ,
\end{equation}
which to second order is equivalent to all the alternative expressions.
It may seem strange to
define parallax, which obviously is an observable quantity, in terms of $b$, which according to
previous discussion is not (directly) observable. However, $\varpi$ should rather be regarded
as a model parameter allowing us to interpret non-barycentric observations in a consistent
manner, and Eq.~(\ref{eq:varpi}) is then the relation to be used in the model for calculating
its effect on the data.

From Eq.~(\ref{eq:mur}) it is seen that the radial proper motion equals $\mu_r=v_r/b$.
This is analogous to the expression for the total (tangential) proper motion
\begin{equation}\label{eq:mut}
\mu \equiv \left(\mu_{\alpha*}^2+\mu_\delta^2\right)^{1/2} = v_t/b
\quad\mbox{or}\quad\vec{\mu}\equiv\vec{v}_t/b \, ,
\end{equation}
where $\vec{v}_t$ is the apparent (or astrometric) tangential velocity (Sect.~\ref{s:light-time}).
The three components of proper motion
$\mu_{\alpha*}$, $\mu_\delta$, and $\mu_r$ are conveniently expressed in the
same unit, for example mas~yr$^{-1}$. The unit of time in this case would be the
Julian year of exactly $365.25\times 86400$~s (TCB).

It is also useful to note the expression for the (apparent) space velocity in terms of the astrometric parameters:
\begin{equation}\label{eq:vel-mu}
  \vec{v}=\frac{A_\mathrm{V}}{\varpi}\left(\vec{r}\mu_r+\vec{\mu}\right)\,,
\end{equation}
where $A_\mathrm{V}=4.740\,470\,446$ equals the astronomical unit expressed in $\mathrm{km}\ \mathrm{yr}\ \mathrm{s}^{-1}$. This relation implies that the parallax and proper motions are expressed in compatible units, for instance, mas and mas~yr$^{-1}$, respectively.

\section{Epoch transformation including light-time
effects}\label{s:including}

In this section, we develop and summarize the transformation of the astrometric parameters and their covariances with rigorous treatment of the light-travel time effects.

\subsection{Solution of the apparent path equation}\label{ss:solution}

Equations (\ref{eq:pos-t-em-t-obs}) and (\ref{eq:apparent-path}) implicitly
determine the apparent position in terms of the true and apparent velocity,
respectively. In subsequent
Eqs.~(\ref{eq:apparent-path-vtrue})--(\ref{eq37})
let $\bar{\vec{v}}$ denote the true velocity.
Using notations introduced in Sect.~\ref{s:astr_params} we can
write Eq.~(\ref{eq:pos-t-em-t-obs}) as
\begin{equation}\label{eq:apparent-path-vtrue}
 \vec{b}=\vec{b}_0+\bar{\vec{v}}\left(t-\frac{b}{c}+\tau_0\right) \, ,
\end{equation}
where $\tau_0=b_0/c$ is the initial light-time. Squaring both sides we
obtain a quadratic equation for the apparent distance:
\begin{equation}\label{eq:quadratic}
 \left(1-\frac{v^2}{c^2}\right)b^2+
 \frac{2}{c}\bar{\vec{v}}'\left[\vec{b}_0+\bar{\vec{v}}\left(t+\tau_0\right)\right]b-
 \left[\vec{b}_0+\bar{\vec{v}}\left(t+\tau_0\right)\right]^2=0 \, .
\end{equation}
It is seen from the Vi{\`e}ta's formulae that the roots of this
equation are of opposite signs. Choosing the positive root, we find
\begin{equation}\label{eq:propagated_distance}
\begin{aligned}
 \left(1-\frac{v^2}{c^2}\right)b=
 &-\frac{1}{c}\bar{\vec{v}}'\left[\vec{b}_0+\bar{\vec{v}}\left(t+\tau_0\right)\right]
\\& +\sqrt{\left[\vec{b}_0+\bar{\vec{v}}\left(t+\tau_0\right)\right]^2-
        \frac{1}{c^2}\left(\bar{\vec{v}}\times\vec{b}_0\right)^2} \, .
\end{aligned}
\end{equation}
The fact that the right-hand side is positive for any position and velocity is easily demonstrated by writing the radicand as the sum of two essentially positive values,
\begin{equation}
    \left(1-\frac{v^2}{c^2}\right)\left[\vec{b}_0+\bar{\vec{v}}\left(t+\tau_0\right)\right]^2+
    \frac{1}{c^2}\left\{\bar{\vec{v}}'\left[\vec{b}_0+\bar{\vec{v}}\left(t+\tau_0\right)\right]\right\}^2\,,
\end{equation}
one of them is exactly equal to the square of the first term in the right-hand side.

Having determined the propagated apparent distance, we can calculate the expression in parenthesis in equation (\ref{eq:apparent-path-vtrue}), which is the difference in the emission time corresponding to the time interval $t$, according to formula (\ref{eq:pos-t-em-t-obs}):
\begin{equation}\label{eq:Tem}
    \Delta T^\mathrm{em}=t-\frac{b}{c}+\tau_0\,.
\end{equation}
Making use of $b$ from Eq. (\ref{eq:propagated_distance}), we get
\begin{equation}
\begin{aligned}
 \left(1-\frac{v^2}{c^2}\right)\Delta T^\mathrm{em}&=
 t+\tau_0+\frac{\bar{\vec{v}}'\vec{b}_0}{c^2}\\
 &-\frac{1}{c}\sqrt{\left[\vec{b}_0+\bar{\vec{v}}\left(t+\tau_0\right)\right]^2-
        \frac{1}{c^2}\left(\bar{\vec{v}}\times\vec{b}_0\right)^2} \, .
\end{aligned}
\end{equation}
It is convenient for the following development to represent
$\Delta T^\mathrm{em}$ as a fraction using the identity
$x-y=(x^2-y^2)/(x+y)$:
\begin{equation}\label{eq37}
 \begin{aligned}
 \Delta&T^\mathrm{em}=
 t\left(t+2\tau_0\right)\times\\
 &\left[t+\tau_0+\dfrac{\bar{\vec{v}}'\vec{b}_0}{c^2}+
  \dfrac{1}{c}
  \sqrt{\left[\vec{b}_0+\bar{\vec{v}} \left(t+\tau_0\right)\right]^2-
        \dfrac{1}{c^2}\left(\bar{\vec{v}}\times\vec{b}_0\right)^2}\,\right]^{-1}.
 \end{aligned}
\end{equation}
Substituting $\bar{\vec{v}}=\vec{v}_0/\left(1-v_{r0}/c\right)$ according to Eq.~(\ref{eq:vel-true-app}) and inserting the resulting expression for $\Delta T^\mathrm{em}$ into Eq.~(\ref{eq:apparent-path}) we finally obtain, after elementary, though rather lengthy calculations,
\begin{equation}\label{eq:apparent-path-solution}
 \vec{b}=\vec{b}_0+\vec{v}_0tf_\mathrm{T}\,,
\end{equation}
where we have introduced the time factor
\begin{equation}
 f_\mathrm{T}=
 \frac{t+2\tau_0}
 {\tau_0+\left(1-\dfrac{v_{r0}}{c}\right)t+
  \dfrac{1}{c}\sqrt{\left(\vec{b}_0+\vec{v}_0t\right)^2+
  \dfrac{2t}{c^2\tau_0}\left(\vec{v}_0\times\vec{b}_0\right)^2}} \, .
\end{equation}
These formulae give the complete solution to the problem: they determine the apparent position at any instant in terms of the given initial apparent position and velocity.

It is instructive to consider briefly the special case of purely radial motion
when $v_{r0}=v_0$ and $v_t=0$. Then, ${\vec{v}_0}'\vec{b}_0=v_0b_0$ and $\vec{v}_0\times\vec{b}_0=0$, so that $f_\mathrm{T}=1$ and
Eq.~(\ref{eq:apparent-path-solution}) simplifies to
\begin{equation}\label{eq:radial-motion}
  \vec{b}=\vec{b}_0+\vec{v}_0t\,.
\end{equation}
Thus, we conclude that the finite light-travel time has no effect on the apparent stellar motion in the case of purely radial motion. Of course, this result can be
obtained much more simply without using the general solution of the apparent path equation: noting that both $\vec{v}^\mathrm{true}$ and $v_r^\mathrm{true}$
are constant in this case, it follows from Eq.~(\ref{eq:vel-app-true}) that the apparent velocity is also constant. Thus, we can replace the differentials in
Eqs.~(\ref{eq:vel-app-true-start}) and (\ref{eq:vel-true-app-start}) with finite differences to find
\begin{equation}
 \Delta T^\mathrm{em}\vec{v}^\mathrm{true}=\Delta T^\mathrm{obs}\vec{v}^\mathrm{app}\,.
\end{equation}
Substituting this in Eq.~(\ref{eq:classical-equation-motion-diff1}) and
using the relation (\ref{eq:true-apparent}) between the true and apparent
positions, we again arrive at Eq.~(\ref{eq:radial-motion}).
Since light-time effects thus vanish in the absence of transverse motion,
they can be expected to be small for stars with small proper motions.
We address this question further in Appendix~\ref{sec:approx}.

\subsection{Propagation of the astrometric parameters}
\label{ss:propagation}

The solution (\ref{eq:apparent-path-solution}) gives the time dependence of the apparent position, which, in turn, determines propagation of the barycentric direction and parallax, defined by Eqs.~(\ref{eq:u}) and (\ref{eq:varpi}), respectively. Squaring both sides of Eq.~(\ref{eq:apparent-path-solution}) we find
\begin{equation}
  b^2=b_0^2\left(1+2\mu_{r0}tf_\mathrm{T}+\left(\mu_0^2+\mu_{r0}^2\right)\left(tf_\mathrm{T}\right)^2\right)\,.
\end{equation}
Here, we used that $\vec{b}'_0\vec{v}_0^{\phantom{2}}=b_0^2\mu_{r0}^{\phantom{2}}$ and $v_0^2=b_0^2\left(\mu_{r0}^2+\mu_0^2\right)$, as is easily seen from Eqs.~(\ref{eq:u}), (\ref{eq:varpi}) and (\ref{eq:vel-mu}).

Introducing the distance factor
\begin{equation}\label{eq:fd}
 f_\mathrm{D}=b_0/b=
\left[1+2\mu_{r0}tf_\mathrm{T}+
 \left(\mu_0^2+\mu_{r0}^2\right)\left(tf_\mathrm{T}\right)^2\right]^{-1/2},
\end{equation}
the propagation of the barycentric direction is
\begin{equation}\label{eq:prop_u}
 \vec{u}=
 \left[\vec{u}_0\left(1+\mu_{r0}tf_\mathrm{T}\right)+
       \vec{\mu}_0tf_\mathrm{T}\right]f_\mathrm{D}
\end{equation}
and the propagation of the parallax becomes
\begin{equation}\label{eq:prop_par}
    \varpi=\varpi_0f_\mathrm{D}\,.
\end{equation}
The celestial coordinates ($\alpha$, $\delta$) at epoch $t$ are obtained from $\vec{u}$ in the usual manner, using Eq.~(\ref{eq:r}).

We now consider the propagation of the proper motions $\vec{\mu}$ and $\mu_r$. It is clear from the above discussion that the proper motions can be found by two equivalent methods. They can be either expressed as the time derivatives or obtained from relevant velocity components. The direct differentiation with respect to time, however, offers great difficulties since $\vec{u}$ and $b$ involve the factor $f_\mathrm{T}$, which
is a complicated function of time. On the contrary, the calculation of the proper motion using the apparent
velocity is relatively simple in the present case.

To find the propagated apparent velocity, it is convenient to employ the following artifice taking advantage of the postulated constancy of the true velocity. We first obtain the true velocity from the initial apparent velocity using Eq.~(\ref{eq:vel-true-app}) and then substitute the true velocity in Eq.~(\ref{eq:vel-app-true}) to get the apparent velocity at the time $t$.
However, it should be emphasized that the radial component of the true velocity $v_r^\mathrm{true}$ in (\ref{eq:vel-app-true}) must be computed
along the propagated barycentric direction:
$v_r^\mathrm{true}=\vec{u}'\vec{v}^\mathrm{true}$. Carrying out the calculation,
we find the propagated apparent velocity
\begin{equation}\label{eq:vel-app-vel-app0}
    \vec{v}=\vec{v}_0 f_\mathrm{V}\,,
\end{equation}
where we have introduced the velocity factor
\begin{equation}\label{eq:fv-definition}
    f_\mathrm{V}=
    \left[1+\dfrac{\vec{u}'\vec{v}_0}{c}-\dfrac{v_{r0}}{c}\right]^{-1}=
    \left[1+\dfrac{(\vec{u}-\vec{u}_0)'\vec{v}_0}{c}\right]^{-1}\,.
\end{equation}
Decomposition of $\vec{v}$ into the components normal and along the propagated direction yields propagation of the transverse and radial components of the apparent velocity, which are
\begin{align}
    \vec{v}_t&=
    \left[\vec{v}_0-\vec{u}\left(\vec{u}'\vec{v}_0\right)\right]
    f_\mathrm{V}\,, \\
    v_r&=
    \left(\vec{u}'\vec{v}_0\right)f_\mathrm{V}\,,
\end{align}
respectively. Substitution of these relations in Eqs.~(\ref{eq:mur}) and (\ref{eq:mut}) gives the propagation of the proper motions:
\begin{align}
    \label{eq:mu_prop}
    \vec{\mu}&=
    \left[\vec{\mu}_0\left(1+\mu_{r0}tf_\mathrm{T}\right)-
    \vec{u}_0\mu_0^2tf_\mathrm{T}\right]
    f_\mathrm{D}^3f_\mathrm{V}\,, \\
    \label{eq:mur_prop}
    \mu_r&=
    \left[\mu_{r0}+\left(\mu_0^2+\mu_{r0}^2\right)tf_\mathrm{T}\right]
    f_\mathrm{D}^2f_\mathrm{V}\,.
\end{align}
To obtain the proper motion components ($\mu_{\alpha*}$, $\mu_\delta$) from vector $\vec{\mu}$ it is necessary to resolve the latter along the tangential vectors $\vec{p}$ and $\vec{q}$, using Eq.~(\ref{eq:proper_motion}). The
tangential vectors are defined in terms of the propagated $\vec{u}$ or ($\alpha$,
$\delta$) at epoch $t$ according to Eq.~(\ref{eq:pq}).

The above formulae describe the complete transformation of ($\alpha_0$, $\delta_0$, $\varpi_0$, $\mu_{\alpha*0}$, $\mu_{\delta 0}$, $\mu_{r0}$) at epoch $T_0$ into ($\alpha$, $\delta$, $\varpi$, $\mu_{\alpha*}$, $\mu_\delta$, $\mu_r$) at the arbitrary epoch $T=T_0+t$, including the light-time effects.
The transformation is rigorously reversible: a second transformation from $T$ to $T_0$ recovers the original six parameters.

\subsection{The scaling factors $f_\mathrm{D}$, $f_\mathrm{T}$, and $f_\mathrm{V}$}
\label{ss:scaling}

The propagation formulae derived in the preceding sections involve three
quantities $f_\mathrm{D}$, $f_\mathrm{T}$, and $f_\mathrm{V}$,
which appear as scaling factors for the changes
in (apparent) distance, time, and velocity over the propagated interval.
In this section, we further examine their physical meaning and
give their expressions in terms of the astrometric parameters.

All three factors are in practice very close to unity, linearly approaching
$1$ as $t\rightarrow 0$. (Approximate formula for small $t$ are
derived in Appendix~\ref{sec:approx}.) While
$f_\mathrm{T}=f_\mathrm{V}=1$ in the limit as $c\rightarrow\infty$,
the distance factor $f_\mathrm{D}$ in general deviates from $1$
when light-time effects are ignored (cf.\ Sect.~\ref{ss:neglecting}),
as it gives the relative change in distance according to Eq.~(\ref{eq:fd}).

The meaning of the time factor $f_\mathrm{T}$ is not immediately evident
from its derivation in Sect.~\ref{ss:solution}. However, noting that it can
also be written
\begin{equation}\label{eq:ft-meaning}
  f_\mathrm{T}=\frac{1}{1-v_{r0}/c}\frac{\Delta T^{\mathrm{em}}}{\Delta T^{\mathrm{obs}}}\,,
\end{equation}
 we see that it represents the combination of two physical effects
 originating from the finiteness of the speed of light: the difference
 between time of observation and time of emission, and the difference
 in the absolute value between the true and apparent velocities (the
 Doppler factor). Writing the initial light-time as
\begin{equation}
  \tau_0=\frac{\tau_\mathrm{A}}{\varpi_0}\,,
\end{equation}
with $\tau_\mathrm{A}=A/c=499.004\,784\,\mathrm s$ being the
light-travel time for the astronomical unit,
we can express the time factor in terms of the astrometric parameters:
\begin{equation}\label{eq:ft}
    f_\mathrm{T}=
    \frac{\varpi_0t+2\tau_\mathrm{A}}
    {\varpi_0t+\tau_\mathrm{A}\left(1+Z-\mu_{r0}t\right)}\,,
\end{equation}
where
\begin{equation}\label{eq:z}
    Z=
    \sqrt{
    1+\left(t+2\tau_\mathrm{A}/\varpi_0\right)\mu_0^2t+\left(2+\mu_{r0}t\right)\mu_{r0}t}\,.
\end{equation}

As shown by Eq.~(\ref{eq:vel-app-vel-app0}), the velocity factor
$f_\mathrm{V}$ yields the relative change in apparent velocity
over the time interval of propagation. From the second equality in
Eq.~(\ref{eq:fv-definition}), this effect can be
understood as a secular change of the Doppler factor. In terms of
the astrometric parameters, the velocity factor can be written as
\begin{equation}
    f_\mathrm{V}=
    \left\{1+\left(\tau_\mathrm{A}/\varpi_0\right)
    \left[\mu_{r0}\left(f_\mathrm{D}-1\right)+f_\mathrm{D}\left(\mu_0^2+\mu_{r0}^2\right)tf_\mathrm{T}
    \right]
    \right\}^{-1}.
\end{equation}

Finally, we note that the factors $f_\mathrm{V}$ and $f_\mathrm{T}$ are
connected by the following remarkably simple relation:
\begin{equation}\label{eq:fv}
    f_\mathrm{V}=
    \frac{\mathrm{d}}{\mathrm{d}t}\left(tf_\mathrm{T}\right)\,.
\end{equation}
While a direct check of this relation involves cumbersome calculations,
it can be verified more easily by comparing the expression for the proper motion
vector in Eq.~(\ref{eq:mu_prop}) with the equivalent vector obtained by
differentiating $\vec{u}$ in Eq.~(\ref{eq:prop_u}) with respect to $t$.
The meaning of Eq.~(\ref{eq:fv}) becomes clearer if it is re-written in the
following way. Substituting Eq.~(\ref{eq:ft-meaning}) for the time factor,
using that $\Delta T^\mathrm{obs}=t$, and making use of
Eqs.~(\ref{eq:vel-app-vel-app0}), (\ref{eq:Tem}), and (\ref{eq:vr-app}),
we obtain the following equivalent form of Eq.~(\ref{eq:fv}):
\begin{equation}
    \frac{v}{v_0}=
    \frac{1-v_{r}/c}{1-v_{r0}/c}\,.
\end{equation}
This equation has a simple interpretation: it gives the explicit relation between the absolute value of the propagated apparent velocity and the propagated
apparent radial velocity. In particular, it shows that these quantities vary in opposite directions: an increasing $v_{r}$ results in a decreasing $v$ and vice versa.
A qualitatively similar behaviour is found in cases when the motion does not
obey the uniform rectilinear model; however, the simple linear relation above does not hold in general.

\subsection{Propagation of errors (covariances)}
\label{sec:propCov}

In this section we consider how uncertainties in the astrometric parameters
$\alpha_0$, $\delta_0$, $\varpi_0$, $\mu_{\alpha*0}$, $\mu_{\delta 0}$, $\mu_{r0}$ at epoch $T_0$
propagate into uncertainties in the transformed parameters
$\alpha$, $\delta$, $\varpi$, $\mu_{\alpha*}$, $\mu_\delta$, $\mu_r$ at epoch $T=T_0+t$.
The uncertainties are quantified by means of the $6\times 6$ covariance matrices $\vec{C}_0$
and $\vec{C}$ in which the rows and columns correspond to the astrometric
parameters taken in the order given above.

The general principle of (linearized) error propagation is well known and we refer to
Appendix~\ref{sec:genErrorProp} for a brief introduction including an illustrative example.
Essentially, it requires the calculation of all 36 partial derivatives constituting the
elements of the Jacobian matrix $\vec{J}$ in Eq.~(\ref{eq:covariance_matrix}), such that
\begin{equation}\label{eq:covariance_matrix1}
    \vec{C}=\vec{J}\vec{C}_0\vec{J}'\,.
\end{equation}
The required partial derivatives are readily found once the relations between the corresponding
differentials have been established, i.e. the first-order propagation of small perturbations of the
parameters. We give below the complete derivation of these differentials since it may be of some
methodological interest. The subsequent determination of the partial derivatives is straightforward,
if somewhat tedious, and the full results are given in Appendix~\ref{app:jacobian_include}.

At this point we need to make one further remark concerning the propagation of perturbations.
The components of proper motion are obtained by resolving the proper motion vector
$\vec{\mu}$ according to Eq.~(\ref{eq:proper_motion}), that is $\mu_{\alpha*}=\vec{p}'\vec{\mu}$
and $\mu_\delta=\vec{q}'\vec{\mu}$. Here, the tangential vectors $\vec{p}$ and $\vec{q}$ are
defined by Eq.~(\ref{eq:pq}) in terms of the barycentric position $(\alpha,\delta)$ of the star at
the relevant epoch. Consider now what happens when both the position and proper motion vectors
receive small perturbations $\Delta\vec{u}$, $\Delta\vec{\mu}$. The perturbation in position
clearly affect $\alpha$ and $\delta$, and the question arises if this also changes $\vec{p}$ and
$\vec{q}$. If that is the case, then the total perturbations on the proper motion components
become
\begin{equation}\label{eq:perturbedMuPQ}
    \Delta\mu_{\alpha*} = \vec{p}'\Delta\vec{\mu} + \Delta\vec{p}'\vec{\mu} \, ,
\quad\quad
    \Delta\mu_\delta = \vec{q}'\Delta\vec{\mu} + \Delta\vec{q}'\vec{\mu} \, ,
\end{equation}
where $\Delta\vec{p}$ and $\Delta\vec{q}$ are the perturbations on the tangential vectors
induced by $\Delta\vec{u}$. The problem here is that the expressions for $\Delta\vec{p}$
and $\Delta\vec{q}$ contain the factors $\sec\delta$ and $\tan\delta$ and therefore may
become arbitrarily large sufficiently close to the celestial poles. In terms of the uncertainties
in $\mu_{\alpha*}$ and $\mu_\delta$ it means that they contain contributions that are
unrelated to the actual uncertainty of the proper motion vector, and which in principle are
unbounded.

Alternatively, it is possible to regard $\vec{p}$ and $\vec{q}$ as a fixed, error-free
reference frame for perturbations in the tangential plane. In this case, we must put
$\Delta\vec{p}=\Delta\vec{p}=\vec{0}$ in Eq.~(\ref{eq:perturbedMuPQ}) and all similar
expressions. This leads to simpler propagation formulae and an intuitively more reasonable
interpretation of the proper motion uncertainties \citep{lindegren1995b}. This option was
adopted for the construction of the Hipparcos and Tycho catalogues
\citep[cf.\ Sect.~1.5.5 in Vol.~1 of][]{esa1997} and is also used in the following.
The practical consequence is that the normal triad $[\vec{p}~\vec{q}~\vec{r}]$ must be
regarded as fixed in the context of perturbations and uncertainties. The triad is conventionally
defined by the adopted values of $(\alpha,\delta)$. This also motivates the formal distinction
between $\vec{r}$ and $\vec{u}$ referred to in footnote~\ref{fn:ru}.

Summarizing in terms of the local coordinate triads, we may say that the calculations below are based on the following postulates:
(i) $\left[\vec{p}_0, \vec{q}_0, \vec{r}_0\right]$ is fixed in space and time and does not depend on the uncertainties of the astrometric parameters;
(ii) $\left[\vec{p}, \vec{q}, \vec{r}\right]$ depends on time through the propagated position $\vec{u}$ as $\vec{r} = \vec{u}$; and
(iii) at any moment of time, $\left[\vec{p}, \vec{q}, \vec{r}\right]$ is fixed in space and does not depend on the uncertainties of the initial astrometric parameters.

Accordingly, if the coordinates receive small perturbations $\Delta{\alpha*}_0$ and $\Delta\delta_0$,
then the perturbed barycentric direction becomes
\begin{equation}
  \vec{u}_0 + \Delta\vec{u}_0=
  \vec{p}_0\Delta{\alpha*}_0+\vec{q}_0\Delta\delta_0
  +\vec{r}_0\left[1-\left(\Delta{\alpha*}_0\right)^2-\left(\Delta\delta_0\right)^2\right]^{1/2}
\end{equation}
where $\Delta{\alpha*}_0=\Delta\delta_0=0$ corresponds to the nominal position.
The quadratic terms follow from the constraint $|{\vec{u}_0}+\Delta\vec{u}_0|=1$.
Taking the time derivative and using the definitions (\ref{eq:mu}) and (\ref{eq:proper_motion}), we have
 \begin{multline}
  \vec{\mu}_0+\Delta\vec{\mu}_0=
  \vec{p}_0\mu_{\alpha*0}+\vec{q}_0\mu_{\delta0}\\
  -\vec{r}_0\left(\mu_{\alpha*0}\Delta{\alpha*}_0+\mu_{\delta0}\Delta\delta_0\right)
  \left[1-\left(\Delta{\alpha*}_0\right)^2-\left(\Delta\delta_0\right)^2\right]^{1/2}.
 \end{multline}
This equation suggests that the full differential of the proper motion vector as a function of the initial coordinates and proper motion components is
\begin{equation}\label{eq:dmu0}
  \mathrm{d}\vec{\mu}_0=-\vec{r}_0\mu_{\alpha*0}\mathrm{d}{\alpha*}_0
  -\vec{r}_0\mu_{\delta0}\mathrm{d}\delta_0
  +\vec{p}_0\mathrm{d}\mu_{\alpha*0}+\vec{q}_0\mathrm{d}\mu_{\delta0}\,.
\end{equation}
The terms in $\mathrm{d}{\alpha*}_0$ and $\mathrm{d}\delta_0$, which are normal to
$\vec{\mu}_0$, lend themselves to a straightforward geometrical interpretation.

%It is worthwhile to make the following remark concerning the calculation of the elements of the Jacobian matrix. According to the general rule, the partial derivatives can be readily found provided that relations between relevant differentials are established. We give below the derivation of the involved differentials since it may be of considerable methodological interest. The determination of the derivatives from known differentials is then relatively straightforward, if somewhat tedious (an example is given at the end of this sections), and the complete expressions are given in Appendix~\ref{app:jacobian_include}.

The propagated astrometric parameters depend on the scaling factors $f_\mathrm{D}$, $f_\mathrm{T}$ and $f_\mathrm{V}$, which in turn are functions of the initial parameters. Thus, the dependence of the propagated parameters on the initial parameters becomes quite involved. To keep the expressions compact, we do not expand the differentials of the scaling factors in what follows and give the complete expressions for them later in this section. Moreover, it is convenient to employ the logarithmic differentials rather than ordinary differentials: $\mathrm{d}\ln f_\mathrm{D}=\mathrm{d}f_\mathrm{D}/f_\mathrm{D}$, etc.

We begin with the differentials of the propagated astrometric parameters. Direct differentiation of Eq.~(\ref{eq:r}) yields
\begin{equation}\label{eq:du_dcoor}
  \mathrm{d}\vec{u}=\vec{p}\mathrm{d}{\alpha*}+\vec{q}\mathrm{d}\delta\,.
\end{equation}
Taking the dot products of this equation with $\vec{p}$ and $\vec{q}$, we get \begin{equation}\label{eq:dcoor_du}
  \mathrm{d}\alpha*=\vec{p}'\mathrm{d}\vec{u}\,,\quad
  \mathrm{d}\delta=\vec{q}'\mathrm{d}\vec{u}\,.
\end{equation}
From Eq.~(\ref{eq:prop_par}) we have
\begin{equation}
  \mathrm{d}\varpi=f_\mathrm{D}\,\mathrm{d}\varpi_0+\varpi\,\mathrm{d}\ln f_\mathrm{D}\,,
\end{equation}
while (\ref{eq:proper_motion}) yields
\begin{equation}\label{eq:dmu_du}
  \mathrm{d}\mu_{\alpha*}=\vec{p}'\mathrm{d}\vec{\mu}\,,\quad
  \mathrm{d}\mu_\delta=\vec{q}'\mathrm{d}\vec{\mu}\,.
\end{equation}
Direct differentiation of Eq.~(\ref{eq:mur_prop}) gives
\begin{equation}\label{eq:dmur}
 \begin{aligned}
    \mathrm{d}\mu_r
    &=
    2tf_\mathrm{T}f_\mathrm{D}^2f_\mathrm{V}
    \left(\mu_{\alpha*0}\mathrm{d}\mu_{\alpha*0}+\mu_{\delta0}\mathrm{d}\mu_{\delta0}\right)\\
    &+
    \left(1+2\mu_{r0}tf_\mathrm{T}\right)f_\mathrm{D}^2f_\mathrm{V}\mathrm{d}\mu_{r0}\\
    &+
    \mu_r
    \left(2\mathrm{d}\ln f_\mathrm{D}+\mathrm{d}\ln f_\mathrm{V}+\mathrm{d}\ln f_\mathrm{T}\right)
    -\mu_{r0}f_\mathrm{D}^2f_\mathrm{V}\mathrm{d}\ln f_\mathrm{T}\,.
 \end{aligned}
\end{equation}

The determination of the differentials of the coordinates and proper motion components from Eqs. (\ref{eq:dcoor_du}) and (\ref{eq:dmu_du}), respectively, requires $\mathrm{d}\vec{u}$ and $\mathrm{d}\vec{\mu}$ to be written in terms of the initial parameters. From Eq.~(\ref{eq:prop_u}) we have
\begin{equation}\label{eq:du}
 \begin{aligned}
    \mathrm{d}\vec{u}=
    &\left[\left(1+\mu_{r0}tf_\mathrm{T}\right)\mathrm{d}\vec{u}_0
          +tf_\mathrm{T}\left(\mathrm{d}\vec{\mu}_0+\vec{r}_0\mathrm{d}\mu_{r0}\right)\right]f_\mathrm{D}\\
    &-\vec{r}_0f_\mathrm{D}\mathrm{d}\ln f_\mathrm{T}
    +\vec{u}\left(\mathrm{d}\ln f_\mathrm{D}+\mathrm{d}\ln f_\mathrm{T}\right)\,.
 \end{aligned}
\end{equation}
It is useful to note that the last term disappears when taking the dot products with $\vec{p}$ and $\vec{q}$, since $\vec{p}'\vec{u}=\vec{q}'\vec{u}=0$; hence it does not contribute to the differentials $\mathrm{d}\alpha*$ and $\mathrm{d}\delta$.
Direct differentiation of Eq.~(\ref{eq:mu_prop}) yields
\begin{equation}\label{eq:dmu}
 \begin{aligned}
    \mathrm{d}\vec{\mu}=
    &-\mu_0^2tf_\mathrm{T}f_\mathrm{D}^3f_\mathrm{V}\mathrm{d}\vec{u}_0
    +
    \left(1+\mu_{r0}tf_\mathrm{T}\right)f_\mathrm{D}^3f_\mathrm{V}\mathrm{d}\vec{\mu}_0\\
    &-2\vec{r}_0tf_\mathrm{T}f_\mathrm{D}^3f_\mathrm{V}
    \left(\mu_{\alpha*0}\mathrm{d}\mu_{\alpha*0}+\mu_{\delta0}\mathrm{d}\mu_{\delta0}\right)\\
    &+\vec{\mu}_0tf_\mathrm{T}f_\mathrm{D}^3f_\mathrm{V}\mathrm{d}\mu_{r0}
    +
    \vec{\mu}
    \left(3\mathrm{d}\ln f_\mathrm{D}+\mathrm{d}\ln f_\mathrm{V}+\mathrm{d}\ln f_\mathrm{T}\right)\\
    &-\vec{\mu}_0f_\mathrm{D}^3f_\mathrm{V}\mathrm{d}\ln f_\mathrm{T}\,.
 \end{aligned}
\end{equation}
 Here we leave the differentials $\mathrm{d}\vec{u}_0$ (analogous to Eq.~\ref{eq:du_dcoor})
 and $\mathrm{d}\vec{\mu}_0$ (from Eq.~\ref{eq:dmu0}) unexpanded to avoid too  lengthy expressions. We use these expressions in the calculation of the partial derivatives listed in Appendix~\ref{app:jacobian_include}.

 Finally, we consider the three scaling factors and start with the distance factor.
 From the definition (\ref{eq:fd}) we find
\begin{equation}
  \begin{aligned}
    \mathrm{d}f_\mathrm{D}=
    &-\left(tf_\mathrm{T}\right)^2f_\mathrm{D}^3
    \left(\mu_{\alpha*0}\mathrm{d}\mu_{\alpha*0}+\mu_{\delta0}\mathrm{d}\mu_{\delta0}\right)\\
    &-tf_\mathrm{T}f_\mathrm{D}^3
    \left(1+\mu_{r0}tf_\mathrm{T}\right)\mathrm{d}\mu_{r0}\\
    &-f_\mathrm{D}^3
    \left[\mu_{r0}t+\left(\mu_0^2+\mu_{r0}^2\right)t^2f_\mathrm{T}\right]\mathrm{d}f_\mathrm{T}\,.
 \end{aligned}
\end{equation}
Making use of the propagated radial proper motion given by Eq. (\ref{eq:mur_prop}), we can write the last term as
$-f_\mathrm{D}\left(\mu_r tf_\mathrm{T}/f_\mathrm{V}\right)\mathrm{d}\ln f_\mathrm{T}$.
Dividing by $f_\mathrm{D}$, we obtain the logarithmic differential
\begin{equation}\label{eq:dlnfd}
 \begin{aligned}
    \mathrm{d}\ln f_\mathrm{D}=
    &-\left(tf_\mathrm{T}\right)^2f_\mathrm{D}^2
    \left(\mu_{\alpha*0}\mathrm{d}\mu_{\alpha*0}+\mu_{\delta0}\mathrm{d}\mu_{\delta0}\right)\\
    &-tf_\mathrm{T}f_\mathrm{D}^2
    \left(1+\mu_{r0}tf_\mathrm{T}\right)\mathrm{d}\mu_{r0}\\
    &-\left(\mu_rtf_\mathrm{T}/f_\mathrm{V}\right)\mathrm{d}\ln f_\mathrm{T}\,.
 \end{aligned}
\end{equation}
The logarithmic differential of the velocity factor is obtained from Eq.~(\ref{eq:fv}) by a straightforward calculation:
\begin{equation}
 \begin{aligned}
    \mathrm{d}\ln f_\mathrm{V}&=
    \frac{\tau_\mathrm{A}}{\varpi_0^2}f_\mathrm{V}
    \left[\mu_{r0}\left(f_\mathrm{D}-1\right)+f_\mathrm{D}\left(\mu_0^2+\mu_{r0}^2\right)tf_\mathrm{T}
    \right]\mathrm{d}\varpi_0\\
    &-2\frac{\tau_\mathrm{A}}{\varpi_0}
    tf_\mathrm{T}f_\mathrm{D}f_\mathrm{V}
    \left(\mu_{\alpha*0}\mathrm{d}\mu_{\alpha*0}+\mu_{\delta0}\mathrm{d}\mu_{\delta0}\right)\\
    &+\frac{\tau_\mathrm{A}}{\varpi_0}
    f_\mathrm{V}
    \left(1-f_\mathrm{D}\left(1+2\mu_{r0}tf_\mathrm{T}\right)\right)\mathrm{d}\mu_{r0}\\
    &-\frac{\tau_\mathrm{A}}{\varpi_0}
    \left(\mu_r/f_\mathrm{D}\right)\mathrm{d}\ln f_\mathrm{D}\\
    &-\frac{\tau_\mathrm{A}}{\varpi_0}f_\mathrm{D}f_\mathrm{V}\left(\mu_0^2+\mu_{r0}^2\right)tf_\mathrm{T}
    \mathrm{d}\ln f_\mathrm{T}.
 \end{aligned}
\end{equation}
Substituting $\mathrm{d}\ln f_\mathrm{D}$ from Eq.~(\ref{eq:dlnfd}), we find
\begin{equation}\label{eq:dlnfv}
 \begin{aligned}
    \mathrm{d}\ln &f_\mathrm{V}=
    \frac{1}{\varpi_0}\left(1-f_\mathrm{V}\right)\mathrm{d}\varpi_0\\
    &+\frac{\tau_\mathrm{A}}{\varpi_0}
    tf_\mathrm{T}f_\mathrm{D}\left(\mu_rtf_\mathrm{T}-2f_\mathrm{V}\right)
    \left(\mu_{\alpha*0}\mathrm{d}\mu_{\alpha*0}+\mu_{\delta0}\mathrm{d}\mu_{\delta0}\right)\\
    &+\frac{\tau_\mathrm{A}}{\varpi_0}
    \left[f_\mathrm{V}
    +f_\mathrm{D}\left(f_\mathrm{V}+\left(1+\mu_{r0}tf_\mathrm{T}\right)\left(\mu_rtf_\mathrm{T}-2f_\mathrm{V}\right)\right)
    \right]\mathrm{d}\mu_{r0}\\
    &-\frac{\tau_\mathrm{A}}{\varpi_0}
    \mu_0^2tf_\mathrm{T}f_\mathrm{D}^3f_\mathrm{V}\mathrm{d}\ln f_\mathrm{T} \, .
 \end{aligned}
\end{equation}
The time factor is conveniently represented as the fraction
\begin{equation}\label{eq:ft_as_fraction}
    f_\mathrm{T}=X/Y
\end{equation}
with $X$ and $Y$ defined by Eq.~(\ref{eq:ft}). Then
\begin{equation}\label{eq:dlnft}
    \mathrm{d}\ln f_\mathrm{T}=\frac{\mathrm{d}X}{X}-\frac{\mathrm{d}Y}{Y}\,,
\end{equation}
where
\begin{equation}\label{eq:dX}
    \mathrm{d}X=t\mathrm{d}\varpi_0
\end{equation}
and
\begin{equation}\label{eq:dY}
 \begin{aligned}
    \mathrm{d}Y=
    &t\left(1-\frac{\mu_0^2\tau_\mathrm{A}^2}{Z\varpi_0^2}\right)\mathrm{d}\varpi_0\\
    &+
    \frac{t\tau_\mathrm{A}}{Z}
    \left(t+2\frac{\tau_\mathrm{A}}{\varpi_0}\right)
    \left(\mu_{\alpha*0}\mathrm{d}\mu_{\alpha*0}+\mu_{\delta0}\mathrm{d}\mu_{\delta0}\right)\\
    &+t\tau_\mathrm{A}\left(\frac{1+\mu_{r0}t}{Z}-1\right)\mathrm{d}\mu_{r0} \, .
 \end{aligned}
\end{equation}

All the required partial derivatives can be derived from the differentials in
the formulae above. As an example, consider the partial derivative of the
propagated right ascension with respect to the initial parallax. From
Eq.~(\ref{eq:dcoor_du}) we have
\begin{equation}
  \frac{\partial\alpha*}{\partial\varpi_0}=\vec{p}'\frac{\partial\vec{u}}{\partial\varpi_0}\,.
\end{equation}
As seen from Eq.~(\ref{eq:du}), the propagated direction $\vec{u}$ depends on $\varpi_0$ only through the time factor, and we can therefore write
\begin{equation}
  \frac{\partial\alpha*}{\partial\varpi_0}=
  -\vec{p}'\vec{r}_0f_\mathrm{D}\frac{\partial\ln f_\mathrm{T}}{\partial\varpi_0}\,.
\end{equation}
The terms in Eqs.~(\ref{eq:dX})--(\ref{eq:dY}) containing $\mathrm{d}\varpi_0$ give
\begin{equation}
  \frac{\partial\ln f_\mathrm{T}}{\partial\varpi_0}=\frac{t}{X}
  -\left(1-\frac{\mu_0^2\tau_\mathrm{A}^2}{Z\varpi_0^2}\right)\frac{t}{Y}\,.
\end{equation}
Writing $X$ and $Y$ in terms of the astrometric parameters, we finally obtain
\begin{equation}
  \frac{\partial\alpha*}{\partial\varpi_0}=
  -\vec{p}'\vec{r}_0f_\mathrm{D}t
  \left[\frac{1}{\varpi_0t+2\tau_\mathrm{A}}
  -\frac{1-\mu_0^2\tau_\mathrm{A}^2/\left(Z\varpi_0^2\right)}{\varpi_0t+\tau_\mathrm{A}\left(1+Z-\mu_{r0}t\right)}
  \right] \,.
\end{equation}

\subsection{Epoch transformation neglecting light-time
effects}\label{ss:neglecting}

Having developed the rigorous formulae including light-time effects, we now consider the case when light-time effects are not important. This formally corresponds to the limit as $c\to\infty$, or zero light-travel time. Since all light-time effects have been parametrized with $\tau_\mathrm{A}=A/c$, we can formally exclude them by putting $\tau_\mathrm{A}=0$. This substitution gives $f_\mathrm{T}=f_\mathrm{V}=1$ and the distance factor
\begin{equation}\label{eq:fd-neglecting-lt}
    f_\mathrm{D}=
    \left[1+2\mu_{r0}t+\left(\mu_0^2+\mu_{r0}^2\right)t^2\right]^{-1/2}\,.
\end{equation}
The propagated quantities are readily obtained: the barycentric direction
\begin{equation}\label{eq:propBarDirection}
    \vec{u}=\left[\vec{r}_0\left(1+\mu_{r0}t\right)+\vec{\mu}_0t\right]f_\mathrm{D}\, ,
\end{equation}
the parallax
\begin{equation}\label{eq:parallax-neglecting-lt}
    \varpi=\varpi_0f_\mathrm{D}\,,
\end{equation}
the proper motion vector
\begin{equation}\label{eq:prop_mu}
    \vec{\mu}=\frac{\mathrm{d}\vec{u}}{\mathrm{d}t}=
    \left[\vec{\mu}_0\left(1+\mu_{r0}t\right)-\vec{r}_0\mu_0^2t\right]f_\mathrm{D}^3\,,
\end{equation}
and the radial proper motion
\begin{equation}\label{eq:prop_mur}
    \mu_r=\frac{\mathrm{d}b}{\mathrm{d}t}\frac{\varpi}{A}=
    \left[\mu_{r0}+\left(\mu_0^2+\mu_{r0}^2\right)t\right]f_\mathrm{D}^2\,.
\end{equation}
The celestial coordinates ($\alpha$, $\delta$) and proper motion components ($\mu_{\alpha*}$, $\mu_\delta$) at epoch $t$ are obtained from $\vec{u}$ and $\vec{\mu}$ in the usual manner using Eqs.~(\ref{eq:r}) and (\ref{eq:proper_motion}), respectively.

These formulae were employed in the reduction procedures used to construct the Hipparcos and Tycho catalogues, since light-time effects were known to be negligible at milli-arcsecond accuracy \citep[][Vol.~1, Sect.~1.5.5]{esa1997}. The transformation described by the above expressions is also rigorously reversible. The corresponding elements of the Jacobian matrix needed to propagate the covariances are given in Appendix~\ref{app:Jacobian_neglect}.

\begin{figure}[t]
\resizebox{\hsize}{!}{\includegraphics{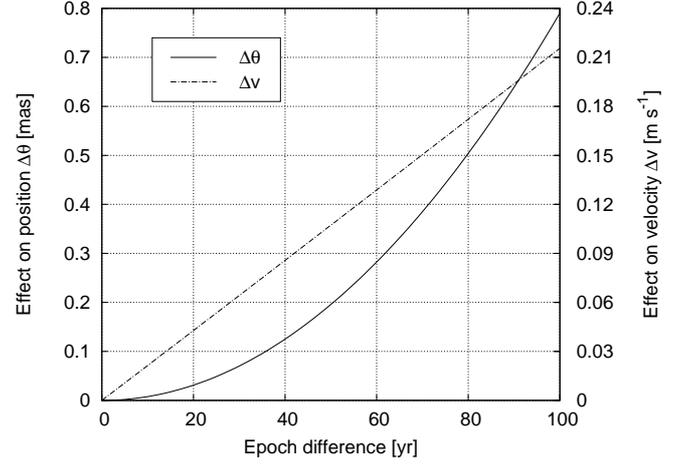}}
% \resizebox{\hsize}{!}{\includegraphics{barnard.pdf}}
  \caption{The effect of light-time on the propagation of the astrometric parameters of
  Barnard's star (HIP~87937). The solid line and left axis show the difference in angular position,
  while the dash-dotted line and right axis show the difference in the the
  apparent space velocity.}
  \label{fig:prop}
\end{figure}

\section{Discussion}\label{s:discussion}

In the following, we discuss the conditions under which the light-time effects
may be significant when propagating the astrometric parameters of a star
from one epoch to another. We consider first the absolute size of the effect
itself, and then its size in relation to deviations from the assumed uniform
rectilinear motion. The applicability of the developed technique to real data
is discussed, and a simple criterion established for when the light-time effects
should be ignored. Finally, we briefly review this work in relation to the earlier
treatment by \citet{stumpff1985}.

\subsection{When is it possible to ignore light-time effects?}\label{ss:possible-to-ignore}

In practice, the finite light-time may be ignored if its observable effects are small
compared to the required astrometric accuracy. In in Appendix~\ref{sec:approx}, we
derive approximate formulae for the effects of the light-time on the parameters
propagated over the time interval $t$. Here, we only consider the effects on the
angular position, $\Delta\theta$, and on the proper motion, $\Delta\mu$.
Let $\sigma_\theta$ and $\sigma_\mu$
be the required accuracies in position and proper motion. The two conditions
for negligible light-time effects are then $\Delta\theta\ll\sigma_\theta$ and
$\Delta\mu\ll\sigma_\mu$, which by means of Eq.~(\ref{eq:effects}) can be
written
\begin{equation}\label{eq:neglect}
  t^2\ll\frac{2\varpi\sigma_\theta}{\mu^3\tau_\mathrm{A}} \, ,
  \quad
  t\ll\frac{\varpi\sigma_\mu}{\mu^3\tau_\mathrm{A}}\,.
\end{equation}
If the positions and proper motions are derived from observations
around the original epoch we find, in the limit of large $t$, that
$\sigma_\theta=t\sigma_\mu$ from Eq.~(\ref{eq:limit-sigma}).
In this case the two conditions become essentially the same (within a
factor of 2).

The strong (cubic) dependence on $\mu$ in Eq.~(\ref{eq:neglect}) suggests
that light-time effects could mainly be important for high-proper motion stars.
As an example, let us consider Barnard's star (HIP~87937) which, with
$\mu=10357.70$~mas~yr$^{-1}$ and $\varpi=549.01$~mas, has the
largest proper motion in the Hipparcos catalogue \citep{esa1997}.
At $\sigma_\theta=1$~mas position accuracy, light-time effects are negligible
for $t\ll 114$~yr. At $\sigma_\theta=1~\mu$as accuracy, they
are only negligible for $t\ll 3.6$~yr. Figure~\ref{fig:prop} shows the effects
in position and velocity ($\Delta v$) as functions of time for this star.

\subsection{When is it necessary to ignore light-time effects?}\label{ss:necessary}

Up until now it has been tacitly assumed  that the astrometric parameters exactly
describe the state of motion of the star. If this condition is not fulfilled, for example,
because of uncertainties in the parameters, a direct application of the simple kinematic model
may produce erroneous and even physically absurd results. Clearly, this will happen for
negative parallaxes, light-time effects must in fact be ignored under more
restrictive conditions.

A simple example illustrates the effect of observational errors. Consider the case when the measured
parallax is smaller than the true parallax, while other astrometric parameters have negligible errors.
The distance inferred from the measured parallax is then too large, and the transverse velocity
calculated from the distance and proper motion is also overestimated. If the true parallax is small,
the measured value, while still positive, can be many times smaller than the true parallax, leading
to distances and transverse velocities overestimated by a large factor. As the observed
parallax goes to zero, the calculated velocity goes to infinity. On the other hand, true velocity must
not exceed the speed of light, $v^\mathrm{true}<c$. Using Eq.~(\ref{eq:vel-true-app}), we find the
condition
\begin{equation}
 \frac{v}{1-v_r/c}<c\,,\quad\mbox{or}\quad v + v_r < c\, ,
\end{equation}
where $v$ and $v_r$ are apparent velocities.
In terms of the astrometric parameters, this gives a constraint on the parallax,
\begin{equation}\label{eq:limitPar}
 \varpi > \tau_\mathrm{A}\left(\mu_r+\sqrt{\mu^2+\mu_r^2}\right)\,.
\end{equation}
For a given proper motion, any parallax below this limit is physically meaningless because
it would correspond to true superluminal motion.%
\footnote{As opposed to an apparent superluminal velocity, $v^\mathrm{app}>c$,
which is physically possible and allowed by Eq.~(\ref{eq:vel-true-app}).}
However, Eq.~(\ref{eq:limitPar}) is a very weak condition: for example, it gives
$\varpi\ga 30~\mu$as for proper motions of the order of 1~arcsec~yr$^{-1}$.
Observational errors put much more stringent constraints on acceptable parallaxes.

\citet{brown1997} discussed the estimation of physical quantities such as stellar distances and
absolute magnitudes from measured trigonometric parallaxes. For individual stars such
estimates are in general significantly biased, unless the ratio of the parallax uncertainty
to the true parallax is less than 0.1. Although this ratio is not precisely known in an actual case,
it may be approximated by the relative error of the measured parallax, $\sigma_\varpi/\varpi$.
Since it can be argued that the application of minute light-time effects is meaningless if the
result is in any case biased by the observational errors, we conclude that light-time effects
should be ignored at least if $\varpi < 10\sigma_\varpi$.

It is worth noting that the propagation formulae obtained by neglecting light-time effects
(i.e. by formally putting $\tau_\mathrm{A}=0$, as in Sect.~\ref{ss:neglecting} and
Appendix~\ref{app:Jacobian_neglect}) work for any value of the parallax. The parallax
only appears in one propagation formula, Eq.~(\ref{eq:parallax-neglecting-lt}), and in the
partial derivatives $J_{34}$, $J_{35}$, and $J_{36}$. All these equations involve the parallax
as a multiplicative factor, creating no formal or numerical problem if the value happens to
be zero or negative. Physically, such a parallax is of course meaningless, but can nevertheless
be regarded as a formal parameter of the model. Brown \citep[in][Ch.~16]{2013asas.book.....V}
gives more general considerations of the use of small, zero, or negative parallaxes in
astrophysical applications of astrometric data.

\begin{table*}
 \caption{Hipparcos stars with significant light-time effect. Columns~1--2 list the
 Hipparcos identifier and parallax, 3--6 the proper motion components,
 including the total (tangential) proper motion $\mu=(\mu_{\alpha*}^2+\mu_\delta^2)^{1/2}$
 and the radial proper motion $\mu_r$, 7--9 the tangential, radial, and total
 velocity, 10 the effect of perspective acceleration over 100~yr, and 11--12
 the light-time effects over 100~yr in position and apparent space velocity.}
 \label{table:1}
 \centering\footnotesize
 \begin{tabular}{ccccccccccccc}
 \hline\hline
 \noalign{\smallskip}
 HIP & $\varpi$ & $\mu_{\alpha*}$ & $\mu_\delta$ & $\mu$ & $\mu_r$         & $v_t$ & $v_r$         & $v$ & $\Delta\theta_\mathrm{persp}$ & $\Delta\theta$ & $\Delta v$ \\
     & (mas)    & (mas yr$^{-1}$) & (mas yr$^{-1}$) & (mas yr$^{-1}$) & (mas yr$^{-1}$) & (km s$^{-1}$) & (km s$^{-1}$) & (km s$^{-1}$) & (mas) & (mas) & (m s$^{-1}$) \\
 \noalign{\smallskip}
 \hline
 \noalign{\smallskip}
\phantom{111}439 &  229.33 & \phantom{1}5634.07 & $-2337.94$ & \phantom{1}6099.89 & \phantom{11}1227.81 & 126.09 & \phantom{11}25.38 &  128.62            & \phantom{1}363 &    0.38 &    0.16 \\
\phantom{11}5336 &  132.40 & \phantom{1}3421.44 & $-1599.27$ & \phantom{1}3776.76 & \phantom{1}$-2739.91$ & 135.22 & \phantom{1}$-98.10$ &  167.06        & \phantom{1}502 &    0.16 &    0.14 \\
\phantom{1}10449 & \phantom{1}16.17 & \phantom{11}994.65 & \phantom{11}$-80.42$ & \phantom{11}997.90 & \phantom{1111}95.82 & 292.55 & \phantom{11}28.09 &  293.89 & \phantom{111}5 &    0.02 &    0.14 \\
\phantom{1}15234 & \phantom{1}30.63 & \phantom{11}739.73 & $-1063.79$ & \phantom{1}1295.70 & \phantom{1}$-1080.99$ & 200.53 & $-167.30$ &  261.15                 & \phantom{11}68 &    0.03 &    0.11 \\
\phantom{1}16209 & \phantom{1}39.19 & \phantom{1}1120.24 & $-1065.81$ & \phantom{1}1546.25 & \phantom{1}$-1430.21$ & 187.04 & $-173.00$ &  254.78                 & \phantom{1}107 &    0.04 &    0.12 \\
 \noalign{\smallskip}
\phantom{1}16404 & \phantom{1}17.58 & \phantom{1}1190.86 & $-1066.16$ & \phantom{1}1598.39 & \phantom{11}$-600.40$ & 431.01 & $-161.90$ &  460.41                 & \phantom{11}47 &    0.09 &    0.51 \\
\phantom{1}18915 & \phantom{1}54.14 & \phantom{1}1732.49 & $-1365.50$ & \phantom{1}2205.93 & \phantom{11}$-294.09$ & 193.15 & \phantom{1}$-25.75$ &  194.86       & \phantom{11}31 &    0.08 &    0.13 \\
\phantom{11}19849\tablefootmark{a} &  198.24 & $-2239.33$ & $-3419.86$ & \phantom{1}4087.79 & \phantom{1}$-1769.76$ & \phantom{1}97.75 & \phantom{1}$-42.32$ &  106.52 & \phantom{1}351 &    0.13 &    0.07 \\
\phantom{1}21609 & \phantom{1}17.00 & \phantom{11}732.93   & \phantom{1}1249.38 &           1448.49 & \phantom{111}211.58   &  403.91 &  \phantom{11}59.00  & 408.20 & \phantom{11}15 &    0.07 &    0.39 \\
\phantom{11}24186\tablefootmark{b} &  255.26 & \phantom{1}6506.05 & $-5731.39$ & \phantom{1}8670.50 & \phantom{1}13202.74 & 161.02 & \phantom{1}245.19 &  293.34 & 5550 &    0.96 &    0.66 \\
 \noalign{\smallskip}
\phantom{1}24316 & \phantom{1}14.55 & \phantom{11}935.43 & \phantom{11}515.36 & \phantom{1}1068.00 & \phantom{111}721.29 & 347.96 & \phantom{1}235.00 &  419.88   & \phantom{11}37 &    0.03 &    0.25 \\
\phantom{1}34285 & \phantom{1}10.68 & \phantom{11}$-93.65$ & \phantom{11}691.23 & \phantom{11}697.55 & \phantom{111}587.27 & 309.62 & \phantom{1}260.67 &  404.74 & \phantom{11}20 &    0.01 &    0.14 \\
\phantom{1}38541 & \phantom{1}35.29 & \phantom{11}705.00 & $-1834.55$ & \phantom{1}1965.35 & \phantom{1}$-1745.34$ & 264.00 & $-234.45$ &  353.08                 & \phantom{1}166 &    0.08 &    0.30 \\
\phantom{1}46120 & \phantom{1}16.46 & \phantom{11}202.13   & \phantom{1}1237.24 &           1253.64 & \phantom{11}$-329.51$ &  361.05 & \phantom{1}$-94.90$ & 373.31 & \phantom{11}20 &    0.05 &    0.27 \\
\phantom{1}48152 & \phantom{1}12.44 & \phantom{11}373.81 & \phantom{1}$-774.75$ & \phantom{11}860.22 & \phantom{111}$-39.36$ & 327.80 & \phantom{1}$-15.00$ &  328.14 & \phantom{111}2 &    0.02 &    0.15 \\
 \noalign{\smallskip}
\phantom{1}49616 & \phantom{11}8.93 & \phantom{1}$-563.56$ & \phantom{11}393.83 & \phantom{1}687.53 & \phantom{11}$-102.29$ &  364.98 & \phantom{1}$-54.30$ & 368.99 & \phantom{111}3 &    0.01 &    0.15 \\
\phantom{1}54035 &  392.40 & \phantom{1}$-580.20$ & $-4767.09$ & \phantom{1}4802.27 & \phantom{1}$-7010.35$ & \phantom{1}58.01 & \phantom{1}$-84.69$ &  102.66 & 1632 &    0.11 &    0.05 \\
\phantom{1}54211 &  206.94 & $-4410.79$ & \phantom{11}943.32 & \phantom{1}4510.53 & \phantom{11}3007.32 & 103.32 & \phantom{11}68.89 &  124.18                & \phantom{1}658 &    0.17 &    0.09 \\
\phantom{1}55042 & \phantom{1}79.71 & $-2465.03$           & \phantom{1}1179.06 &           2732.50 & \phantom{11}$-588.52$ &  162.51 & \phantom{1}$-35.00$ & 166.23 & \phantom{11}78 &    0.10 &    0.12 \\
\phantom{1}56936 & \phantom{1}44.28 & \phantom{11}262.62 & $-3157.21$ & \phantom{1}3168.11 & \phantom{1}$-1102.22$ & 339.17 & $-118.00$ &  359.11                     & \phantom{1}169 &    0.28 &    0.62 \\
 \noalign{\smallskip}
\phantom{11}57939\tablefootmark{c} &  109.21 & \phantom{1}4003.69 & $-5813.00$ & \phantom{1}7058.36 & \phantom{1}$-2265.77$ & 306.38 & \phantom{1}$-98.35$ &  321.78            & \phantom{1}775 &    1.24 &    1.13 \\
\phantom{11}74234\tablefootmark{d} & \phantom{1}33.68 & $-1001.47$ & $-3542.66$ & \phantom{1}3681.49 & \phantom{11}2207.95 & 518.17 &  \phantom{1}310.77 &  604.22    & \phantom{1}394 &    0.57 &    1.86 \\
\phantom{11}74235\tablefootmark{d} & \phantom{1}34.14 & \phantom{1}$-998.86$ & $-3542.91$ & \phantom{1}3681.02 & \phantom{11}2233.43 & 511.12 & \phantom{1}310.12 &  597.85 & \phantom{1}399 &    0.56 &    1.82 \\
\phantom{1}76976 & \phantom{1}17.44 & $-1115.54$ & \phantom{1}$-302.77$ & \phantom{1}1155.90 & \phantom{11}$-629.54$ & 314.19 & $-171.12$ &  357.77                   & \phantom{11}35 &    0.03 &    0.21 \\
\phantom{1}80837 & \phantom{1}24.34 & \phantom{1}$-432.73$ & $-1392.34$ & \phantom{1}1458.03 & \phantom{11}$-244.56$ & 283.97 & \phantom{1}$-47.63$ &  287.93         & \phantom{11}17 &    0.05 &    0.19 \\
 \noalign{\smallskip}
\phantom{11}87937\tablefootmark{e} &  549.01 & \phantom{1}$-797.84$ & 10326.93 & 10357.70 & $-12798.54$ & \phantom{1}89.43 & $-110.51$ &  142.17               & 6427 &    0.79 &    0.21 \\
100568 & \phantom{1}22.88 & \phantom{11}539.73 & $-1055.93$ & \phantom{1}1185.87 & \phantom{11}$-827.36$ & 245.70 & $-171.42$ &  299.59                               & \phantom{11}48 &    0.03 &    0.14 \\
104059 & \phantom{1}52.26 & \phantom{1}$-914.54$ &          $-2035.65$&           2231.65 & \phantom{111}$-15.43$  &  202.43 & \phantom{11}$-1.40$ & 202.44 & \phantom{111}2 &    0.08 &    0.15 \\
\phantom{1}104214\tablefootmark{f} &  287.13 & \phantom{1}4155.10 & \phantom{1}3258.90 & \phantom{1}5280.65 & \phantom{1}$-3981.87$ & \phantom{1}87.18 & \phantom{1}$-65.74$ &  109.19 & 1019 &    0.20 &    0.08 \\
\phantom{1}104217\tablefootmark{f} &  285.42 & \phantom{1}4107.40 & \phantom{1}3143.72 & \phantom{1}5172.40 & \phantom{1}$-3857.60$ & \phantom{1}85.91 & \phantom{1}$-64.07$ &  107.17 & \phantom{1}967 &    0.19 &    0.08 \\
 \noalign{\smallskip}
108870 &  275.76 & \phantom{1}3959.97 & $-2538.84$ & \phantom{1}4703.94 & \phantom{1}$-2326.86$ & \phantom{1}80.86 & \phantom{1}$-40.00$ & \phantom{1}90.22    & \phantom{1}531 &    0.15 &    0.06 \\
114046 &  303.90 & \phantom{1}6767.26 & \phantom{1}1326.66 & \phantom{1}6896.07 & \phantom{111}564.79 & 107.57 & \phantom{111}8.81 &  107.93                  & \phantom{1}189 &    0.41 &    0.13 \\
117254 & \phantom{1}12.91 & \phantom{11}255.64 & \phantom{1}$-856.78$ & \phantom{11}894.11 & \phantom{111}$-29.68$ & 328.31 & \phantom{1}$-10.90$ &  328.49           & \phantom{111}1 &    0.02 &    0.16 \\
 \noalign{\smallskip}
\hline
 \end{tabular}
 \tablefoot{
  \tablefoottext{a}{40 Eri.}
  \tablefoottext{b}{Kapteyn's star.}
  \tablefoottext{c}{Groombridge 1830 (Argelander's star).}
  \tablefoottext{d}{Component of the quadruple star system CCDM J15102-1624.}
  \tablefoottext{e}{Barnard's star.}
  \tablefoottext{f}{Component of the double star 61 Cyg.}
 }
\end{table*}

\subsection{Light-time effects for Hipparcos stars}\label{ss:for-Hipparcos}

For a star with known parallax, proper motion, and radial velocity the
effects of the finite light-time on the propagated astrometric parameters
are readily computed from a direct comparison between the rigorous
propagation (Sect.~\ref{ss:propagation}) and when
light-time effects are ignored (Sect.~\ref{ss:neglecting}). We have made this
computation for Hipparcos stars with radial velocities taken from the XHIP
catalogue \citep{anderson+francis2012}. The last two columns in
Table~\ref{table:1} list the computed effects over a century in position
($\Delta\theta$) and in the absolute value of the space velocity ($\Delta v$).
For the listed 33 entries, these quantities exceed 0.1~mas or 0.1~m~s$^{-1}$,
respectively. Since $\Delta\theta$ increases very nearly quadratically
with time, and $\Delta v$ increases linearly (cf.\ Fig.~\ref{fig:prop}),
the values listed in the table for $t=100$~yr are easily scaled to other
epoch differences.

To find the objects listed in Table~\ref{table:1} we computed the effects
for all 15\,517 objects with a radial velocity in XHIP and $\varpi>10\sigma_\varpi$ in the Hipparcos catalogue \citep{esa1997}.
It should be noted that the Hipparcos catalogue may contain additional
entries where the effects exceed 0.1~mas or 0.1~m~s$^{-1}$, but
which were excluded because of one of the above criteria.

It is interesting to compare the light-time effect in position with the more well-known
perspective acceleration \citep{schlesinger1917,vanDeKamp1977,murray1983},
which is a purely geometrical effect due to the changing distance and angle
between the velocity vector and line of sight (it is equivalent to the Coriolis acceleration
in the coordinate system rotating with the line of sight). The apparent acceleration
due to this effect is $\dot{\mu}=-2\mu\mu_r$, which after time $t$ results
in the positional offset
\begin{equation}\label{eq:persp}
  \Delta\theta_\mathrm{persp} = \mu\mu_r t^2
\end{equation}
(dropping the negative sign).
As can be seen from the table, the perspective effect is typically some
three orders of magnitude greater than the light-time effect (roughly the ratio of the
speed of light to the stellar velocity). However, the actual ratio depends on the
angle between the stellar motion and the line of sight, so that for example the star
with the largest light-time effect in position (HIP~57939) has only the sixth largest perspective
effect. The perspective acceleration is fully taken into
account in all the propagation formulae presented in this paper, both with and
without the light-time effects.

\begin{figure}
\resizebox{\hsize}{!}{\includegraphics{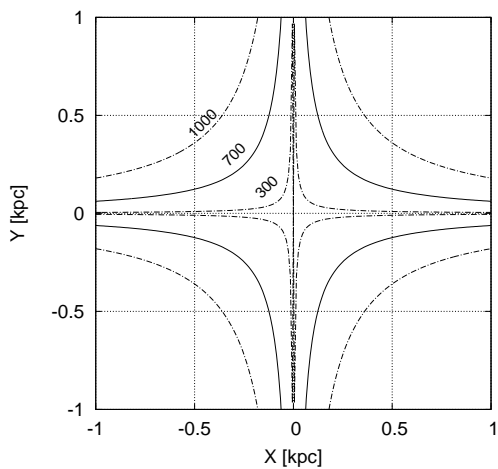}}
% \resizebox{\hsize}{!}{\includegraphics{comparison.pdf}}
  \caption{Comparison of the effects of the light-time and Galactic acceleration on the
  position in the Galactic plane. The Sun is situated at $x=y=0$, with the Galactic centre
  on the $x$ axis. The curves show where the two effects are equal for the tangential
  velocity indicated (in km~s$^{-1}$) next to the curve. The light-time effect exceeds that
  of the Galactic acceleration for stars closer to the Sun than the curves. Beyond the curves,
  the effect of the Galactic acceleration is more significant.}
  \label{fig:comparison}
\end{figure}

\subsection{Light-time effects and the Galactic acceleration}\label{ss:galacc}

The propagation formulae are based on the uniform rectilinear model of stellar motions,
which at some point breaks down due to the (differential) Galactic acceleration. This effect
is estimated in Appendix~\ref{ss:model_applicability}, where Eq.~(\ref{eq:delta-theta}) is
derived for the positional offset caused by the acceleration in the Galactic plane. This
offset is proportional to $\sin 2\glon$, where $\glon$ is Galactic longitude, and increases
quadratically with time. Since the light-time effect in position also increases quadratically,
as given by Eq.~(\ref{eq:effects_physical}), the relative size of the two effects do not change
with time and depend only on the position and velocity of the star in question. Comparing the
two formulae it is seen that the light-time effect dominates over the Galactic acceleration if
\begin{equation}\label{eq:equal-effects}
    0.36
    \times\left(\frac{v_t}{10^3~\mathrm{km~s}^{-1}}\right)^3
    >
    \left(\frac{b}{1~\mathrm{kpc}}\right)^{2}
    \times\bigl|\sin 2\glon\bigr|\,.
\end{equation}
In a rectangular, heliocentric Galactic coordinate system with $x$ and $y$ axes
directed towards $\glon=0\degr$ and $90\degr$, respectively, we find
$\sin 2\glon=2xy/(x^2+y^2)=2xy/b^2$. For a given transverse velocity $v_t$,
the light-time effect therefore dominates inside an area around the Sun delimited
by four hyperbolas, one in each quadrant, as shown in Fig.~\ref{fig:comparison}.
Since very few Galactic stars have velocities exceeding 300~km~s$^{-1}$, we
conclude that the effect of the Galactic acceleration generally dominates beyond
a distance of about 100~pc from the Sun.

\subsection{Relation to the work by \citet{stumpff1985}}\label{ss:stumpff}

The most complete analysis of the astrometric light-time effects prior to the present
treatment was the pioneering study by \citet{stumpff1985}. As in the present work, Stumpff
carried out his analysis within the framework of the uniform rectilinear model, made an
explicit distinction between true and apparent quantities, solved the quadratic light-time
equation, and expressed the propagated quantities as functions of the time of observation.
However, there are a number of important differences between the two studies,
summarized hereafter.

\citet{stumpff1985} expressed the propagated quantities in terms of the true, not apparent,
velocity. Although he derived the transformation from true to apparent velocity equivalent
to our Eq.~(\ref{eq:vel-app-true}), he did not give the inverse relation in
Eq.~(\ref{eq:vel-true-app}). Instead, an approximate equation for the inverse transformation
was derived via the relativistic formula for the Doppler effect. In our opinion, this obscures
the treatment by mixing in a quite different problem, namely the relation between the
(astrometric) radial velocity $v_r$ and the spectroscopically observable Doppler effect
\citep{lindegren_dravins2003}.

Concerning the transformation of the astrometric parameters from one epoch to another,
\cite{stumpff1985} proposed an iterative method to find the true parameters, then propagate
them, and finally recover the apparent parameters at the new epoch. By contrast, we give the propagated
apparent parameters in closed form as functions of the initial apparent parameters.

Furthermore, Stumpff developed the propagation formulae in terms of the arc length along
the apparent stellar path, i.e. essentially in scalar form, while we use vectors throughout.
The vector formalism yields clear and concise formulae for the explicit transformations,
which are readily translated to computer code. This is obviously important for any practical
application of the model, where also the error propagation needs to be considered. Clearly,
Stumpff did not provide the Jacobian of the transformation, nor did he consider the
limitations of the uniform rectilinear model in the Galactic potential.

\section{Conclusions}\label{s:conclusion}

We have presented a technique for transforming astrometric data from one epoch to another based on the uniform rectilinear model of barycentric stellar motion, including a rigorous treatment of the effects of light travel time.

A consistent treatment of light-time necessitates distinguishing between true and apparent (observed) position and velocity. While the former are in principle unknown, they may nevertheless be inferred from an assumed model of stellar motion. The six astrometric parameters (two components of the position, the trigonometric parallax, and three components of the proper motion) are defined with respect to the apparent quantities. Applying the light-time equation to the uniform motion model, we derived the path equation in terms of the apparent position and velocity.

The analytical solution of the apparent path equation for uniform rectilinear motion gives the propagated barycentric position at any instance of time. The postulated constancy of the true velocity enables us to find the propagated apparent velocity. Remarkably, all the light-time effects are conveniently parametrized by two factors, which are equal to 1 when light-time is ignored. We obtain explicit formulae for the propagated astrometric parameters both in the case when light-time effects are included and when they are neglected. We also provide the corresponding elements of the Jacobian matrix to be used in the propagation of covariances. Thus, we have derived a complete set of formulae for the rigorous and fully reversible propagation of astrometric data, and their covariances, over arbitrary time intervals.

The effect of the light-time on the astrometric parameters is roughly proportional to $\mu^3$. Although the light-time effects are generally very small, they are significant for high-velocity stars within a few tens of pc from the Sun, where they exceed the effects of the curvature of their orbits in the Galactic potential.

Distance should be well known to allow meaningful calculation of the light-time effects. Therefore, the epoch propagation including light-time should only be used for stars with reliable parallax distances:
we recommend the criterion $\varpi > 10\sigma_\varpi$. Astrometric parameters of stars with smaller parallaxes should be propagated neglecting light-time. Thus, the presented technique is
applicable both to high- and low-accuracy astrometric data provided that the proper mode of epoch transformation is selected.

\begin{acknowledgements}
LL gratefully acknowledges support from the Swedish National Space Board. AGB is grateful to Lund Observatory for their warm hospitality during his short-term visits. AGB also acknowledges the support from the Deutsche Zentrum f\"ur Luft- und Raumfahrt e.V. (DLR). We warmly thank our referee, Anthony G. A. Brown (Leiden Observatory), for his valuable comments and suggestions.
\end{acknowledgements}

\bibliographystyle{aa}
\bibliography{bl}

\appendix

\section{General error propagation}
\label{sec:genErrorProp}

Although the propagation of errors is discussed in many textbooks \citep[see, for example,][]{brandt,bevington+robinson}, we find it instructive for the subsequent discussion to give a brief exposition of this technique.

In the context of the error propagation, it is convenient to represent the astrometric parameters by a vector $\vec{a}$ of length 6. All information on the standard errors, $\sigma$, of the parameters and correlations between them is contained in the $6\times 6$ variance-covariance matrix $\vec{C}$ with the elements of the latter being
\begin{equation}
    c_{ii}=\sigma_i^2\quad c_{ik}=\rho_{ik}\sigma_i\sigma_k\,,
\end{equation}
where $\rho_{ik}$ is the correlation coefficient of $i$-th and $k$-th parameter.

If vector of the parameters $\vec{a}_0$ undergoes a transformation giving new vector,  $\vec{a}=\vec{f}\left(\vec{a}_0\right)$, small variations in the parameters are related as
\begin{equation}
 \Delta a_i=\sum_k\frac{\partial f_i}{\partial a_k}\Delta a_{0k}\,.
\end{equation}
In matrix form this can be written:
\begin{equation}\label{eq:Delta_a}
 \Delta\vec{a}=\vec{J}\Delta\vec{a}_0\,,
\end{equation}
where $\vec{J}$ is the Jacobian matrix of the transformation:
\begin{equation}
 J_{ik}=\frac{\partial f_i}{\partial a_k}
\end{equation}
evaluated at the point $\vec{a}_0$. Now let $\Delta\vec{a}$ be the difference between the estimated and true parameter vectors. If the estimate is unbiased, then $\mathrm E\left(\Delta\vec{a}_0\right)=\vec{0}$, where $\mathrm E$ is the expectation operator, and the covariance matrix of $\vec{a}_0$ is given by $\vec{C}_0=\mathrm E\left(\Delta\vec{a}_0\Delta\vec{a}_0'\right)$, with the prime denoting matrix transposition. It follows from (\ref{eq:Delta_a}) that $\vec{a}$ is also unbiased, to the first order in the errors, and that its covariance is given by
\begin{equation}\label{eq:covariance_matrix}
    \vec{C}=\vec{J}\vec{C}_0\vec{J}'\,.
\end{equation}
This equation is the basis for the error propagation discussed below.

If the inverse function $\vec{f}^{-1}$ exists, then it is possible to transform the data set $\left[\vec{a}, \vec{C}\right]$ back to the original form $\left[\vec{a}_0, \vec{C}_0\right]$, and the two representations can be regarded as equivalent from the point of view of information content. A necessary condition for this is that $\left|\vec{J_f}\right|\neq 0$,
in which case ${\vec{J}}_{\vec{f}^{-1}}=\left(\vec{J}_{\vec{f}}\right)^{-1}$. The transformations discussed here satisfy this condition.

\paragraph{A simple example:}

To illustrate the general error propagation technique using the Jacobian, we give below some very
simplified formulae. We emphasize that they should not be used for actual calculations,
but are only given for illustration.
The simplistic formulae for transforming a celestial position $\left(\alpha,\delta\right)$ over the epoch
difference $t$ are
\begin{equation}\label{eq:simplified}
 \begin{aligned}
  \alpha &=\alpha_0+t\mu_{\alpha0}\,,\\
  \delta &=\delta_0+t\mu_{\delta0}\,.
 \end{aligned}
\end{equation}
It is useful to point out that in this equation the proper motion in right ascension does not contain
the factor $\cos\delta$. This is not a good physical model of how the stars move on the sky: in
general it describes a curved, spiralling motion towards one of the poles, whereas real (unperturbed)
stars are expected to move along great-circle arcs. Although the difference with respect to the
rigorous model (Sect.~\ref{s:including}) is often very small, it becomes significant over long time
intervals or for stars near the celestial poles. In this model, the changes in the proper motion
components and in the parallax are neglected and the Jacobian matrix for the epoch transformation
is then:
\begin{equation}\label{eq:jacob}
 \vec{J}=
 \left(
 \begin{array}{cccccc}
  1 & 0 & 0 & t & 0 & 0 \\
  0 & 1 & 0 & 0 & t & 0 \\
  0 & 0 & 1 & 0 & 0 & 0 \\
  0 & 0 & 0 & 1 & 0 & 0 \\
  0 & 0 & 0 & 0 & 1 & 0 \\
  0 & 0 & 0 & 0 & 0 & 1
 \end{array}
 \right)\,.
\end{equation}
The inverse transformation is obtained by reversing the sign of $t$. It is
easily verified that the resulting matrix is indeed the inverse of Eq.~(\ref{eq:jacob}).

The covariance matrix for the six astrometric parameters at
epoch $t$ are obtained from (\ref{eq:covariance_matrix}); this
yields in particular for the variances in position:
\begin{equation}
 \begin{aligned}
  \sigma_{\alpha}^2 &=
   \left[
    \sigma_{\alpha}^2
   +2t\rho_{\alpha}^{\mu_{\alpha}}
    \sigma_{\alpha}\sigma_{\mu_{\alpha}}
   +t^2\sigma_{\mu_{\alpha}}^2
   \right]_0\,,\\
  \sigma_{\delta}^2 &=
   \left[
    \sigma_{\delta}^2
   +2t\rho_{\delta}^{\mu_{\delta}}
    \sigma_{\delta}\sigma_{\mu_{\delta}}
   +t^2\sigma_{\mu_{\delta}}^2
   \right]_0\\
 \end{aligned}
\end{equation}
(with all quantities in the right members referring to the initial epoch).
Here the notation $\rho_x^y$ means the coefficient of correlation between the
astrometric parameters $x$ and $y$.

Finally, let us consider an extreme case of very large epoch difference. Putting formally $t\to\infty$, we find that
\begin{equation}\label{eq:limit-sigma}
 \sigma_{\alpha}\to t\sigma_{\mu_{\alpha}}\,,\quad\sigma_{\delta}\to t\sigma_{\mu_{\delta}}\,,
\end{equation}
while $\sigma_\varpi$, $\sigma_{\mu_\alpha}$, $\sigma_{\mu_\delta}$, and $\sigma_{\mu_r}$
are unchanged.
Similarly, after direct calculations, we obtain the limiting forms of all nine correlation coefficients affected by the transformation:
\begin{multline}\label{eq:limit-rho}
  \rho_{\alpha}^\delta\to\rho_{\mu_{\alpha}}^{\mu_{\delta}}\,,~~
  \rho_{\alpha}^\varpi\to\rho_\varpi^{\mu_{\alpha}}\,,~~
  \rho_{\alpha}^{\mu_{\alpha}}\to1\,,~~
  \rho_{\alpha}^{\mu_{\delta}}\to\rho_{\mu_{\alpha}}^{\mu_{\delta}}\,,~~
  \rho_{\alpha}^{\mu_r}\to\rho_{\mu_{\alpha}}^{\mu_r}\,,\\
  \rho_\delta^\varpi\to\rho_\varpi^{\mu_{\delta}}\,,~~
  \rho_\delta^{\mu_{\alpha*}}\to\rho_{\mu_{\alpha*}}^{\mu_{\delta}}\,,~~
  \rho_\delta^{\mu_{\delta}}\to1\,,~~
  \rho_\delta^{\mu_r}\to\rho_{\mu_{\delta}}^{\mu_r}\,.
\end{multline}
Although the terms in the right-hand sides of Eqs.~(\ref{eq:limit-sigma}) and (\ref{eq:limit-rho}) refer to the initial epoch, we do not show it explicitly because these quantities remain unchanged.

Thus all information about initial covariances of the positions becomes less significant as the epoch difference increases and vanishes in the long run. Similar arguments hold for the rigorous propagation, except that they cannot be demonstrated so easily.

\paragraph{Initialization of $\vec{C}_0$:}

The initial covariance matrix $\vec{C}_0$ must be specified in order to calculate the covariance matrix
of the propagated astrometric parameters $\vec{C}$. Available astrometric catalogues seldom give the
correlations between the parameters, nor do they usually contain radial velocities. Absence of the
correlations does not create any problems for the error propagation since all the off-diagonal elements of
$\vec{C}_0$ are just set to zero, but the radial velocity is crucial for the rigorous propagation.
While the Hipparcos and Tycho catalogues provide the complete first five rows and columns of
$\vec{C}_0$, this matrix must therefore be augmented with a sixth row and column related to the initial
radial proper motion $\mu_{r0}$. If the initial radial velocity $v_{r0}$ has the standard error
$\sigma_{v_{r0}}$ and is assumed to be statistically independent of the astrometric parameters
in the catalogue, then the required additional elements in $\vec{C}_0$ are
\begin{equation}
 \begin{aligned}
  \left[C_0\right]_{i6}=\left[C_0\right]_{6i}=\left[C_0\right]_{i3}\left(v_{r0}/A\right),
  \quad i=1\dots5\,,\\
  \left[C_0\right]_{66}=\left[C_0\right]_{33}\left(v_{r0}^2+\sigma_{v_{r0}}^2\right)/A^2
  +\left(\varpi_0\sigma_{v_{r0}}/A\right)^2\,
 \end{aligned}
\end{equation}
\citep{esa1997,michalik+2014}.
If the radial velocity is not known, it is recommended that $v_{r0}=0$ is used, together with an
appropriately large value of $\sigma_{v_{r0}}$ (set to, for example, the expected velocity dispersion of the
stellar type in question), in which case $\left[C_0\right]_{66}$ in general is still
positive. This means
that the unknown perspective acceleration is accounted for in the uncertainty of the
propagated astrometric parameters.

It should be noted that strict reversal of the transformation (from $T$ to $T_0$), according
to the standard model of stellar motion, is only possible if the full six-dimensional
parameter vector and covariance is considered.

\section{Elements of the Jacobian matrix}\label{app:jacobian_include}

This appendix gives explicit formulae for the 36 partial derivatives constituting the Jacobian matrix
needed to calculate the covariance matrix of the propagated parameters according to
Eq.~(\ref{eq:covariance_matrix}).
In what follows, we introduce symbols $\chi$ to designate the partial derivatives of the logarithm of the
velocity factor:
\begin{equation*}
  \mathrm{d}\ln f_\mathrm{V}=\chi_\varpi\mathrm{d}\varpi_0
  +\chi_\mu\left(\mu_{\alpha*0}\mathrm{d}\mu_{\alpha*0}+\mu_{\delta0}\mathrm{d}\mu_{\delta0}\right)
  +\chi_r\mathrm{d}\mu_{r0}+\chi_{\mathrm{T}}\mathrm{d}\ln f_{\mathrm{T}}\,.
\end{equation*}
It can be seen from Eq. (\ref{eq:dlnfv}) that
\begin{equation*}
 \begin{aligned}
  \chi_\varpi&=\frac{1}{\varpi_0}\left(1-f_\mathrm{V}\right)\,,\\
  \chi_\mu&=\frac{\tau_\mathrm{A}}{\varpi_0}
    tf_\mathrm{T}f_\mathrm{D}\left(\mu_rtf_\mathrm{T}-2f_\mathrm{V}\right)\,,\\
  \chi_r&=\frac{\tau_\mathrm{A}}{\varpi_0}
  \left[f_\mathrm{V}
    +f_\mathrm{D}\left(f_\mathrm{V}+\left(1+\mu_{r0}tf_\mathrm{T}\right)\left(\mu_rtf_\mathrm{T}-2f_\mathrm{V}\right)\right)
    \right]\,,\\
  \chi_{\mathrm{T}}&=-\frac{\tau_\mathrm{A}}{\varpi_0}f_\mathrm{D}^3
    \mu_0^2tf_\mathrm{T}\,.
 \end{aligned}
\end{equation*}
Similarly, the logarithmic differential of the time factor can be written as
\begin{equation*}
  \mathrm{d}\ln f_\mathrm{T}=
  \psi_\varpi\mathrm{d}\varpi_0
  +\psi_\mu\left(\mu_{\alpha*0}\mathrm{d}\mu_{\alpha*0}+\mu_{\delta0}\mathrm{d}\mu_{\delta0}\right)
  +\psi_r\mathrm{d}\mu_{r0}\,,
\end{equation*}
where, as it follows from Eqs. (\ref{eq:dlnft}), (\ref{eq:dX}) and (\ref{eq:dY}),
\begin{equation*}
 \begin{aligned}
  \psi_\varpi&=\frac{t}{X}-\frac{t}{Y}\left(1-\frac{\mu_0^2\tau_\mathrm{A}^2}{Z\varpi_0^2}\right)\,,\\
  \psi_\mu&=-\frac{t\tau_\mathrm{A}}{YZ}\left(t+2\frac{\tau_\mathrm{A}}{\varpi_0}\right)\,,\\
  \psi_r&=-\frac{t\tau_\mathrm{A}}{Y}\left(\frac{1+\mu_{r0}t}{Z}-1\right)\,.
 \end{aligned}
\end{equation*}
 The quantities $X$, $Y$, and $Z$ are defined by the Eqs. (\ref{eq:ft_as_fraction}), (\ref{eq:ft}) and (\ref{eq:z}). We give below, for reference, these quantities explicitly:
\begin{equation*}
 \begin{aligned}
  X&=\varpi_0t+2\tau_\mathrm{A}\,,\\
  Y&=\varpi_0t+\tau_\mathrm{A}\left(1+Z-\mu_{r0}t\right)\,,\\
  Z&=\sqrt{
    1+\left(t+2\tau_\mathrm{A}/\varpi_0\right)\mu_0^2t+\left(2+\mu_{r0}t\right)\mu_{r0}t}\,.
 \end{aligned}
\end{equation*}
It is convenient to eliminate $\mathrm{d}\ln f_\mathrm{D}$ and $\mathrm{d}\ln f_\mathrm{V}$ from the expressions for the differentials of the proper motions (\ref{eq:dmur}) and (\ref{eq:dmu}), replacing them by $\mathrm{d}\ln f_\mathrm{T}$ and the differentials of the astrometric parameters. To simplify following formulae, we introduce special designations for the coefficients of $\mathrm{d}\ln f_\mathrm{T}$ in $\mathrm{d}\vec{\mu}$ and $\mathrm{d}\mu_r$, respectively:
\begin{equation*}
 \begin{aligned}
  \vec{\nu}&=\vec{\mu}\left[
    1-tf_\mathrm{T}
    \left(3\frac{\mu_r}{f_\mathrm{V}}
    +\frac{\tau_\mathrm{A}}{\varpi_0}\mu_0^2f_\mathrm{D}^3f_\mathrm{V}
    \right)
    \right]
    -\vec{\mu}_0f_\mathrm{D}^3f_\mathrm{V}\,,\\
  \xi&=\mu_r\left[
    1-tf_\mathrm{T}
    \left(2\frac{\mu_r}{f_\mathrm{V}}
    +\frac{\tau_\mathrm{A}}{\varpi_0}\mu_0^2f_\mathrm{D}^3f_\mathrm{V}
    \right)
    \right]
    -\mu_{r0}f_\mathrm{D}^2f_\mathrm{V}\,.\\
 \end{aligned}
\end{equation*}

We, moreover, show how the partial derivatives of the propagated positions with respect to the initial radial proper motion may be expressed in terms of the propagated proper motions. As it has been noted, the term proportional to the propagated barycentric position $\vec{u}$ in Eq.~(\ref{eq:du}) is not significant because it is normal to both $\vec{p}$ and $\vec{q}$. However, keeping the first item in this term, $\vec{u}\,\mathrm{d}\ln f_\mathrm{D}$, and using the Eq.~(\ref{eq:dlnfd}) for $\mathrm{d}\ln f_\mathrm{D}$, we can write the derivative as
\begin{equation*}
 \frac{\partial\vec{u}}{\partial\mu_{r0}}=\left[\vec{r}_0f_\mathrm D-\vec{u}\left(1+\mu_{r0}tf_\mathrm T\right)f_\mathrm D^2\right]tf_\mathrm T-\vec{r}_0f_\mathrm D\psi_r\,.
\end{equation*}
Substituting Eq.~(\ref{eq:prop_u}) for $\vec{u}$ and making use of the propagation of the proper motion given by Eq.~(\ref{eq:prop_mu}), we find that
\begin{equation*}
 \frac{\partial\vec{u}}{\partial\mu_{r0}}=-\vec{\mu}\left(tf_\mathrm T\right)^2/f_\mathrm V-\vec{r}_0f_\mathrm D\psi_r\,.
\end{equation*}
Taking the dot products with $\vec{p}$ and $\vec{q}$, we finally get the formulae for $J_{16}$ and $J_{26}$ given below, respectively.

The elements of the Jacobian matrix are given hereafter.
\begin{equation*}
 \begin{aligned}
    J_{11}=\frac{\partial\alpha*}{\partial{\alpha*}_0}=
    \vec{p}'\vec{p}_0\left(1+\mu_{r0}tf_\mathrm{T}\right)f_\mathrm{D}
    -\vec{p}'\vec{r}_0\mu_{\alpha*0}tf_\mathrm{T}f_\mathrm{D}
 \end{aligned}
\end{equation*}
\begin{equation*}
 \begin{aligned}
    J_{12}=\frac{\partial\alpha*}{\partial\delta_0}=
    \vec{p}'\vec{q}_0\left(1+\mu_{r0}tf_\mathrm{T}\right)f_\mathrm{D}
    -\vec{p}'\vec{r}_0\mu_{\delta0}tf_\mathrm{T}f_\mathrm{D}
 \end{aligned}
\end{equation*}
\begin{equation*}
 \begin{aligned}
    J_{13}=\frac{\partial\alpha*}{\partial\varpi_0}=
    -\vec{p}'\vec{r}_0f_\mathrm{D}\psi_\varpi
 \end{aligned}
\end{equation*}
\begin{equation*}
 \begin{aligned}
    J_{14}=\frac{\partial\alpha*}{\partial\mu_{\alpha*0}}=
    \vec{p}'\vec{p}_0tf_\mathrm{T}f_\mathrm{D}
    -\vec{p}'\vec{r}_0\mu_{\alpha*0}f_\mathrm{D}\psi_\mu
 \end{aligned}
\end{equation*}
\begin{equation*}
 \begin{aligned}
    J_{15}=\frac{\partial\alpha*}{\partial\mu_{\delta0}}=
    \vec{p}'\vec{q}_0tf_\mathrm{T}f_\mathrm{D}
    -\vec{p}'\vec{r}_0\mu_{\delta0}f_\mathrm{D}\psi_\mu
 \end{aligned}
\end{equation*}
\begin{equation*}
 \begin{aligned}
    J_{16}=\frac{\partial\alpha*}{\partial\mu_{r0}}=
    -\mu_{\alpha*}\left(tf_\mathrm{T}\right)^2/f_\mathrm v
    -\vec{p}'\vec{r}_0f_\mathrm{D}\psi_r
 \end{aligned}
\end{equation*}
\begin{equation*}
 \begin{aligned}
    J_{21}=\frac{\partial\delta}{\partial{\alpha*}_0}=
    \vec{q}'\vec{p}_0\left(1+\mu_{r0}tf_\mathrm{T}\right)f_\mathrm{D}
    -\vec{q}'\vec{r}_0\mu_{\alpha*0}tf_\mathrm{T}f_\mathrm{D}
 \end{aligned}
\end{equation*}
\begin{equation*}
 \begin{aligned}
    J_{22}=\frac{\partial\delta}{\partial\delta_0}=
    \vec{q}'\vec{q}_0\left(1+\mu_{r0}tf_\mathrm{T}\right)f_\mathrm{D}
    -\vec{q}'\vec{r}_0\mu_{\delta0}tf_\mathrm{T}f_\mathrm{D}
 \end{aligned}
\end{equation*}
\begin{equation*}
 \begin{aligned}
    J_{23}=\frac{\partial\delta}{\partial\varpi_0}=
    -\vec{q}'\vec{r}_0f_\mathrm{D}\psi_\varpi
 \end{aligned}
\end{equation*}
\begin{equation*}
 \begin{aligned}
    J_{24}=\frac{\partial\delta}{\partial\mu_{\alpha*0}}=
    \vec{q}'\vec{p}_0tf_\mathrm{T}f_\mathrm{D}
    -\vec{q}'\vec{r}_0\mu_{\alpha*0}f_\mathrm{D}\psi_\mu
 \end{aligned}
\end{equation*}
\begin{equation*}
 \begin{aligned}
    J_{25}=\frac{\partial\delta}{\partial\mu_{\delta0}}=
    \vec{q}'\vec{q}_0tf_\mathrm{T}f_\mathrm{D}
    -\vec{q}'\vec{r}_0\mu_{\delta0}f_\mathrm{D}\psi_\mu
 \end{aligned}
\end{equation*}
\begin{equation*}
 \begin{aligned}
    J_{26}=\frac{\partial\delta}{\partial\mu_{r0}}=
    -\mu_{\delta}\left(tf_\mathrm{T}\right)^2/f_\mathrm V
    -\vec{q}'\vec{r}_0f_\mathrm{D}\psi_r
 \end{aligned}
\end{equation*}
\begin{equation*}
 \begin{aligned}
    J_{31}=\frac{\partial\varpi}{\partial{\alpha*}_0}=0
 \end{aligned}
\end{equation*}
\begin{equation*}
 \begin{aligned}
    J_{32}=\frac{\partial\varpi}{\partial\delta_0}=0
 \end{aligned}
\end{equation*}
\begin{equation*}
 \begin{aligned}
    J_{33}=\frac{\partial\varpi}{\partial\varpi_0}=
    f_\mathrm{D}
    -\varpi\left(\mu_rtf_\mathrm{T}/f_\mathrm{V}\right)\psi_\varpi
 \end{aligned}
\end{equation*}
\begin{equation*}
 \begin{aligned}
    J_{34}=\frac{\partial\varpi}{\partial\mu_{\alpha*0}}=
    -\varpi\mu_{\alpha*0}\left(tf_\mathrm{T}\right)^2f_\mathrm{D}^2
    -\varpi\mu_{\alpha*0}\left(\mu_rtf_\mathrm{T}/f_\mathrm{V}\right)\psi_\mu
 \end{aligned}
\end{equation*}
\begin{equation*}
 \begin{aligned}
    J_{35}=\frac{\partial\varpi}{\partial\mu_{\delta0}}=
    -\varpi\mu_{\delta0}\left(tf_\mathrm{T}\right)^2f_\mathrm{D}^2
    -\varpi\mu_{\delta0}\left(\mu_rtf_\mathrm{T}/f_\mathrm{V}\right)\psi_\mu
 \end{aligned}
\end{equation*}
\begin{equation*}
 \begin{aligned}
    J_{36}=\frac{\partial\varpi}{\partial\mu_{r0}}=
    -\varpi\left(1+\mu_{r0}tf_\mathrm{T}\right)tf_\mathrm{T}f_\mathrm{D}^2
    -\varpi\left(\mu_rtf_\mathrm{T}/f_\mathrm{V}\right)\psi_r
 \end{aligned}
\end{equation*}
\begin{equation*}
 \begin{aligned}
    J_{41}=\frac{\partial\mu_{\alpha*}}{\partial{\alpha*}_0}=
    -\vec{p}'\vec{p}_0\mu_0^2tf_\mathrm{T}f_\mathrm{D}^3f_\mathrm{V}
    -\vec{p}'\vec{r}_0\mu_{\alpha*0}\left(1+\mu_{r0}tf_\mathrm{T}\right)f_\mathrm{D}^3f_\mathrm{V}
 \end{aligned}
\end{equation*}
\begin{equation*}
 \begin{aligned}
    J_{42}=\frac{\partial\mu_{\alpha*}}{\partial\delta_0}=
    -\vec{p}'\vec{q}_0\mu_0^2tf_\mathrm{T}f_\mathrm{D}^3f_\mathrm{V}
    -\vec{p}'\vec{r}_0\mu_{\delta0}\left(1+\mu_{r0}tf_\mathrm{T}\right)f_\mathrm{D}^3f_\mathrm{V}
 \end{aligned}
\end{equation*}
\begin{equation*}
 \begin{aligned}
    J_{43}=\frac{\partial\mu_{\alpha*}}{\partial\varpi_0}=
    \vec{p}'\vec{\nu}\psi_\varpi
 \end{aligned}
\end{equation*}
\begin{equation*}
 \begin{aligned}
    J_{44}=\frac{\partial\mu_{\alpha*}}{\partial\mu_{\alpha*0}} &=
    \vec{p}'\vec{p}_0\left(1+\mu_{r0}tf_\mathrm{T}\right)f_\mathrm{D}^3f_\mathrm{V}
    -2\vec{p}'\vec{r}_0\mu_{\alpha*0}tf_\mathrm{T}f_\mathrm{D}^3f_\mathrm{V}\\
    &-3\mu_{\alpha*}\mu_{\alpha*0}\left(tf_\mathrm{T}\right)^2f_\mathrm{D}^2f_\mathrm{V}
    +\mu_{\alpha*}\mu_{\alpha*0}\chi_\mu
    +\vec{p}'\vec{\nu}\mu_{\alpha*0}\psi_\mu
 \end{aligned}
\end{equation*}
\begin{equation*}
 \begin{aligned}
    J_{45}=\frac{\partial\mu_{\alpha*}}{\partial\mu_{\delta0}} &=
    \vec{p}'\vec{q}_0\left(1+\mu_{r0}tf_\mathrm{T}\right)f_\mathrm{D}^3f_\mathrm{V}
    -2\vec{p}'\vec{r}_0\mu_{\delta0}tf_\mathrm{T}f_\mathrm{D}^3f_\mathrm{V}\\
    &-3\mu_{\alpha*}\mu_{\delta0}\left(tf_\mathrm{T}\right)^2f_\mathrm{D}^2f_\mathrm{V}
    +\mu_{\alpha*}\mu_{\delta0}\chi_\mu
    +\vec{p}'\vec{\nu}\mu_{\delta0}\psi_\mu
 \end{aligned}
\end{equation*}
\begin{equation*}
 \begin{aligned}
    J_{46}=\frac{\partial\mu_{\alpha*}}{\partial\mu_{r0}}&=
    \vec{p}'\left[\vec{\mu}_0f_\mathrm{D}-3\vec{\mu}\left(1+\mu_{r0}tf_\mathrm{T}\right)\right]
    tf_\mathrm{T}f_\mathrm{D}^2f_\mathrm{V}\\
    &+\mu_{\alpha*}\chi_r+\vec{p}'\vec{\nu}\psi_r
 \end{aligned}
\end{equation*}
\begin{equation*}
 \begin{aligned}
    J_{51}=\frac{\partial\mu_\delta}{\partial{\alpha*}_0}=
    -\vec{q}'\vec{p}_0\mu_0^2tf_\mathrm{T}f_\mathrm{D}^3f_\mathrm{V}
    -\vec{q}'\vec{r}_0\mu_{\alpha*0}\left(1+\mu_{r0}tf_\mathrm{T}\right)f_\mathrm{D}^3f_\mathrm{V}
 \end{aligned}
\end{equation*}
\begin{equation*}
 \begin{aligned}
    J_{52}=\frac{\partial\mu_\delta}{\partial\delta_0}=
    -\vec{q}'\vec{q}_0\mu_0^2tf_\mathrm{T}f_\mathrm{D}^3f_\mathrm{V}
    -\vec{q}'\vec{r}_0\mu_{\delta0}\left(1+\mu_{r0}tf_\mathrm{T}\right)f_\mathrm{D}^3f_\mathrm{V}
 \end{aligned}
\end{equation*}
\begin{equation*}
 \begin{aligned}
    J_{53}=\frac{\partial\mu_\delta}{\partial\varpi_0}=
    \vec{q}'\vec{\nu}\psi_\varpi
 \end{aligned}
\end{equation*}
\begin{equation*}
 \begin{aligned}
    J_{54}=\frac{\partial\mu_\delta}{\partial\mu_{\alpha*0}} &=
    \vec{q}'\vec{p}_0\left(1+\mu_{r0}tf_\mathrm{T}\right)f_\mathrm{D}^3f_\mathrm{V}
    -2\vec{q}'\vec{r}_0\mu_{\alpha*0}tf_\mathrm{T}f_\mathrm{D}^3f_\mathrm{V}\\
    &-3\mu_\delta\mu_{\alpha*0}\left(tf_\mathrm{T}\right)^2f_\mathrm{D}^2f_\mathrm{V}
    +\mu_{\delta}\mu_{\alpha*0}\chi_\mu
    +\vec{q}'\vec{\nu}\mu_{\alpha*0}\psi_\mu
 \end{aligned}
\end{equation*}
\begin{equation*}
 \begin{aligned}
    J_{55}=\frac{\partial\mu_\delta}{\partial\mu_{\delta0}} &=
    \vec{q}'\vec{q}_0\left(1+\mu_{r0}tf_\mathrm{T}\right)f_\mathrm{D}^3f_\mathrm{V}
    -2\vec{q}'\vec{r}_0\mu_{\delta0}tf_\mathrm{T}f_\mathrm{D}^3f_\mathrm{V}\\
    &-3\mu_\delta\mu_{\delta0}\left(tf_\mathrm{T}\right)^2f_\mathrm{D}^2f_\mathrm{V}
    +\mu_{\delta}\mu_{\delta0}\chi_\mu
    +\vec{q}'\vec{\nu}\mu_{\delta0}\psi_\mu
 \end{aligned}
\end{equation*}
\begin{equation*}
 \begin{aligned}
    J_{56}=\frac{\partial\mu_\delta}{\partial\mu_{r0}}&=
    \vec{q}'\left[\vec{\mu}_0f_\mathrm{D}-3\vec{\mu}\left(1+\mu_{r0}tf_\mathrm{T}\right)\right]
    tf_\mathrm{T}f_\mathrm{D}^2f_\mathrm{V}\\
    &+\mu_{\delta}\chi_r
    +\vec{q}'\vec{\nu}\psi_r
 \end{aligned}
\end{equation*}
\begin{equation*}
 \begin{aligned}
    J_{61}=\frac{\partial\mu_r}{\partial{\alpha*}_0}=0
 \end{aligned}
\end{equation*}
\begin{equation*}
 \begin{aligned}
    J_{62}=\frac{\partial\mu_r}{\partial\delta_0}=0
 \end{aligned}
\end{equation*}
\begin{equation*}
 \begin{aligned}
    J_{63}=\frac{\partial\mu_r}{\partial\varpi_0}=\xi\psi_\varpi
 \end{aligned}
\end{equation*}
\begin{equation*}
 \begin{aligned}
    J_{64}=\frac{\partial\mu_r}{\partial\mu_{\alpha*0}}=
    2\mu_{\alpha*0}\left(1+\mu_{r0}tf_\mathrm{T}\right)tf_\mathrm{T}f_\mathrm{D}^4f_\mathrm{V}
    +\mu_{\alpha*0}\mu_r\chi_\mu
    +\mu_{\alpha*0}\xi\psi_\mu
 \end{aligned}
\end{equation*}
\begin{equation*}
 \begin{aligned}
    J_{65}=\frac{\partial\mu_r}{\partial\mu_{\delta0}}=
    2\mu_{\delta0}\left(1+\mu_{r0}tf_\mathrm{T}\right)tf_\mathrm{T}f_\mathrm{D}^4f_\mathrm{V}
    +\mu_{\delta0}\mu_r\chi_\mu
    +\mu_{\delta0}\xi\psi_\mu
 \end{aligned}
\end{equation*}
\begin{equation*}
 \begin{aligned}
    J_{66}=\frac{\partial\mu_r}{\partial\mu_{r0}}=
    \left[\left(1+\mu_{r0}tf_\mathrm{T}\right)^2-\mu_0^2\left(tf_\mathrm{T}\right)^2\right]
    f_\mathrm{D}^4f_\mathrm{V}
    +\mu_r\chi_r
    +\xi\psi_r
 \end{aligned}
\end{equation*}

\section{Elements of the Jacobian matrix neglecting light-time effects}\label{app:Jacobian_neglect}

This appendix gives explicit formulae for the 36 partial derivatives constituting the Jacobian matrix
of the propagated astrometric parameters for the case when
light-time effects are not taken into account. The following formulae can be obtained either by a direct
differentiation of the corresponding equations in Sect.~\ref{ss:neglecting}, or more easily by putting
$f_\mathrm{T}=f_\mathrm{V}=1$ and $\tau_\mathrm{A}=0$ in the derivatives in
Appendix~\ref{app:jacobian_include}. The elements given below are equivalent to the elements
given in Vol.~1, Sect.~1.5.5 of the Hipparcos and Tycho catalogues \citep{esa1997}. In that
publication, the radial proper motion $\mu_r$ is denoted $\zeta$, and the distance factor
$f_\mathrm{D}$ is denoted $f$.
\begin{equation*}
 \begin{aligned}
    J_{11}=\frac{\partial\alpha*}{\partial{\alpha*}_0}=
    \vec{p}'\vec{p}_0\left(1+\mu_{r0}t\right)f_\mathrm{D}
    -\vec{p}'\vec{r}_0\mu_{\alpha*0}tf_\mathrm{D}
 \end{aligned}
\end{equation*}
\begin{equation*}
 \begin{aligned}
    J_{12}=\frac{\partial\alpha*}{\partial\delta_0}=
    \vec{p}'\vec{q}_0\left(1+\mu_{r0}t\right)f_\mathrm{D}
    -\vec{p}'\vec{r}_0\mu_{\delta0}tf_\mathrm{D}
 \end{aligned}
\end{equation*}
\begin{equation*}
 \begin{aligned}
    J_{13}=\frac{\partial\alpha*}{\partial\varpi_0}=0
 \end{aligned}
\end{equation*}
\begin{equation*}
 \begin{aligned}
    J_{14}=\frac{\partial\alpha*}{\partial\mu_{\alpha*0}}=
    \vec{p}'\vec{p}_0tf_\mathrm{D}
 \end{aligned}
\end{equation*}
\begin{equation*}
 \begin{aligned}
    J_{15}=\frac{\partial\alpha*}{\partial\mu_{\delta0}}=
    \vec{p}'\vec{q}_0tf_\mathrm{D}
 \end{aligned}
\end{equation*}
\begin{equation*}
 \begin{aligned}
    J_{16}=\frac{\partial\alpha*}{\partial\mu_{r0}}=
    -\mu_{\alpha*}t^2
 \end{aligned}
\end{equation*}
\begin{equation*}
 \begin{aligned}
    J_{21}=\frac{\partial\delta}{\partial{\alpha*}_0}=
    \vec{q}'\vec{p}_0\left(1+\mu_{r0}t\right)f_\mathrm{D}
    -\vec{q}'\vec{r}_0\mu_{\alpha*0}tf_\mathrm{D}
 \end{aligned}
\end{equation*}
\begin{equation*}
 \begin{aligned}
    J_{22}=\frac{\partial\delta}{\partial\delta_0}=
    \vec{q}'\vec{q}_0\left(1+\mu_{r0}t\right)f_\mathrm{D}
    -\vec{q}'\vec{r}_0\mu_{\delta0}tf_\mathrm{D}
 \end{aligned}
\end{equation*}
\begin{equation*}
 \begin{aligned}
    J_{23}=\frac{\partial\delta}{\partial\varpi_0}=0
 \end{aligned}
\end{equation*}
\begin{equation*}
 \begin{aligned}
    J_{24}=\frac{\partial\delta}{\partial\mu_{\alpha*0}}=
    \vec{q}'\vec{p}_0tf_\mathrm{D}
 \end{aligned}
\end{equation*}
\begin{equation*}
 \begin{aligned}
    J_{25}=\frac{\partial\delta}{\partial\mu_{\delta0}}=
    \vec{q}'\vec{q}_0tf_\mathrm{D}
 \end{aligned}
\end{equation*}
\begin{equation*}
 \begin{aligned}
    J_{26}=\frac{\partial\delta}{\partial\mu_{r0}}=
    -\mu_{\delta}t^2
 \end{aligned}
\end{equation*}
\begin{equation*}
 \begin{aligned}
    J_{31}=\frac{\partial\varpi}{\partial{\alpha*}_0}=0
 \end{aligned}
\end{equation*}
\begin{equation*}
 \begin{aligned}
    J_{32}=\frac{\partial\varpi}{\partial\delta_0}=0
 \end{aligned}
\end{equation*}
\begin{equation*}
 \begin{aligned}
    J_{33}=\frac{\partial\varpi}{\partial\varpi_0}=f_\mathrm{D}
 \end{aligned}
\end{equation*}
\begin{equation*}
 \begin{aligned}
    J_{34}=\frac{\partial\varpi}{\partial\mu_{\alpha*0}}=
    -\varpi\mu_{\alpha*0}t^2f_\mathrm{D}^2
 \end{aligned}
\end{equation*}
\begin{equation*}
 \begin{aligned}
    J_{35}=\frac{\partial\varpi}{\partial\mu_{\delta0}}=
    -\varpi\mu_{\delta0}t^2f_\mathrm{D}^2
 \end{aligned}
\end{equation*}
\begin{equation*}
 \begin{aligned}
    J_{36}=\frac{\partial\varpi}{\partial\mu_{r0}}=
    -\varpi\left(1+\mu_{r0}t\right)tf_\mathrm{D}^2
 \end{aligned}
\end{equation*}
\begin{equation*}
 \begin{aligned}
    J_{41}=\frac{\partial\mu_{\alpha*}}{\partial{\alpha*}_0}=
    -\vec{p}'\vec{p}_0\mu_0^2tf_\mathrm{D}^3
    -\vec{p}'\vec{r}_0\mu_{\alpha*0}\left(1+\mu_{r0}t\right)f_\mathrm{D}^3
 \end{aligned}
\end{equation*}
\begin{equation*}
 \begin{aligned}
    J_{42}=\frac{\partial\mu_{\alpha*}}{\partial\delta_0}=
    -\vec{p}'\vec{q}_0\mu_0^2tf_\mathrm{D}^3
    -\vec{p}'\vec{r}_0\mu_{\delta0}\left(1+\mu_{r0}t\right)f_\mathrm{D}^3
 \end{aligned}
\end{equation*}
\begin{equation*}
 \begin{aligned}
    J_{43}=\frac{\partial\mu_{\alpha*}}{\partial\varpi_0}=0
 \end{aligned}
\end{equation*}
\begin{equation*}
 \begin{aligned}
    J_{44}=\frac{\partial\mu_{\alpha*}}{\partial\mu_{\alpha*0}} =
    \vec{p}'\vec{p}_0\left(1+\mu_{r0}t\right)f_\mathrm{D}^3
    -2\vec{p}'\vec{r}_0\mu_{\alpha*0}tf_\mathrm{D}^3
    -3\mu_{\alpha*}\mu_{\alpha*0}t^2f_\mathrm{D}^2 \\
 \end{aligned}
\end{equation*}
\begin{equation*}
 \begin{aligned}
    J_{45}=\frac{\partial\mu_{\alpha*}}{\partial\mu_{\delta0}} =
    \vec{p}'\vec{q}_0\left(1+\mu_{r0}t\right)f_\mathrm{D}^3
    -2\vec{p}'\vec{r}_0\mu_{\delta0}tf_\mathrm{D}^3
    -3\mu_{\alpha*}\mu_{\delta0}t^2f_\mathrm{D}^2 \\
 \end{aligned}
\end{equation*}
\begin{equation*}
 \begin{aligned}
    J_{46}=\frac{\partial\mu_{\alpha*}}{\partial\mu_{r0}}=
    \vec{p}'\left[\vec{\mu}_0f_\mathrm{D}-3\vec{\mu}\left(1+\mu_{r0}t\right)\right]tf_\mathrm{D}^2
 \end{aligned}
\end{equation*}
\begin{equation*}
 \begin{aligned}
    J_{51}=\frac{\partial\mu_\delta}{\partial{\alpha*}_0}=
    -\vec{q}'\vec{p}_0\mu_0^2tf_\mathrm{D}^3
    -\vec{q}'\vec{r}_0\mu_{\alpha*0}\left(1+\mu_{r0}t\right)f_\mathrm{D}^3
 \end{aligned}
\end{equation*}
\begin{equation*}
 \begin{aligned}
    J_{52}=\frac{\partial\mu_\delta}{\partial\delta_0}=
    -\vec{q}'\vec{q}_0\mu_0^2tf_\mathrm{D}^3
    -\vec{q}'\vec{r}_0\mu_{\delta0}\left(1+\mu_{r0}t\right)f_\mathrm{D}^3
 \end{aligned}
\end{equation*}
\begin{equation*}
 \begin{aligned}
    J_{53}=\frac{\partial\mu_\delta}{\partial\varpi_0}=0
 \end{aligned}
\end{equation*}
\begin{equation*}
 \begin{aligned}
    J_{54}=\frac{\partial\mu_\delta}{\partial\mu_{\alpha*0}} =
    \vec{q}'\vec{p}_0\left(1+\mu_{r0}t\right)f_\mathrm{D}^3
    -2\vec{q}'\vec{r}_0\mu_{\alpha*0}tf_\mathrm{D}^3
    -3\mu_\delta\mu_{\alpha*0}t^2f_\mathrm{D}^2 \\
 \end{aligned}
\end{equation*}
\begin{equation*}
 \begin{aligned}
    J_{55}=\frac{\partial\mu_\delta}{\partial\mu_{\delta0}} =
    \vec{q}'\vec{q}_0\left(1+\mu_{r0}t\right)f_\mathrm{D}^3
    -2\vec{q}'\vec{r}_0\mu_{\delta0}tf_\mathrm{D}^3
    -3\mu_\delta\mu_{\delta0}t^2f_\mathrm{D}^2 \\
 \end{aligned}
\end{equation*}
\begin{equation*}
 \begin{aligned}
    J_{56}=\frac{\partial\mu_\delta}{\partial\mu_{r0}}=
    \vec{q}'\left[\vec{\mu}_0f_\mathrm{D}-3\vec{\mu}\left(1+\mu_{r0}t\right)\right]tf_\mathrm{D}^2
 \end{aligned}
\end{equation*}
\begin{equation*}
 \begin{aligned}
    J_{61}=\frac{\partial\mu_r}{\partial{\alpha*}_0}=0
 \end{aligned}
\end{equation*}
\begin{equation*}
 \begin{aligned}
    J_{62}=\frac{\partial\mu_r}{\partial\delta_0}=0
 \end{aligned}
\end{equation*}
\begin{equation*}
 \begin{aligned}
    J_{63}=\frac{\partial\mu_r}{\partial\varpi_0}=0
 \end{aligned}
\end{equation*}
\begin{equation*}
 \begin{aligned}
    J_{64}=\frac{\partial\mu_r}{\partial\mu_{\alpha*0}}=
    2\mu_{\alpha*0}\left(1+\mu_{r0}t\right)tf_\mathrm{D}^4
 \end{aligned}
\end{equation*}
\begin{equation*}
 \begin{aligned}
    J_{65}=\frac{\partial\mu_r}{\partial\mu_{\delta0}}=
    2\mu_{\delta0}\left(1+\mu_{r0}t\right)tf_\mathrm{D}^4
 \end{aligned}
\end{equation*}
\begin{equation*}
 \begin{aligned}
    J_{66}=\frac{\partial\mu_r}{\partial\mu_{r0}}=
    \left[\left(1+\mu_{r0}t\right)^2-\mu_0^2t^2\right]f_\mathrm{D}^4
 \end{aligned}
\end{equation*}

\section{Approximate formulae for the light-time effects}
\label{sec:approx}

In this appendix we derive approximate formulae for the effects of the light-time on
the propagated astrometric parameters. These formulae should not be used for the actual
propagation, but only to estimate the significance of the effects.

It is clear from Sect.~\ref{ss:scaling} that the light-time effects are
determined by the scaling factors in time and velocity, $f_\mathrm{T}$ and $f_\mathrm{V}$.
Since these factors are very
close to unity, it is useful to introduce two small quantities, $\varepsilon_\mathrm{T}$ and
$\varepsilon_\mathrm{V}$, which can be regarded as small parameters of the employed formalism:
\begin{equation}\label{eq:ftv-epsilon}
  f_\mathrm{T}=1+\varepsilon_\mathrm{T}\quad\mbox{and}\quad f_\mathrm{V}=1+\varepsilon_\mathrm{V}\,.
\end{equation}
Since $\varepsilon_\mathrm{T}$ and $\varepsilon_\mathrm{V}$ are zero at $t=0$, it is convenient to represent them as an explicit functions of time. Expanding Eqs. (\ref{eq:ft}) and (\ref{eq:fv}) in a Taylor series in time and keeping the first-order terms, we find that
\begin{equation}
  \varepsilon_\mathrm{T}=-\frac{\mu_0^2\tau_\mathrm{A}}{2\varpi_0}t
  \quad\mbox{and}\quad
  \varepsilon_\mathrm{V}=-\frac{\mu_0^2\tau_\mathrm{A}}{\varpi_0}t\,,
\end{equation}
i. e. $\varepsilon_\mathrm{V}=2\varepsilon_\mathrm{T}$.

As the next step, we express the propagated astrometric parameters as linear functions of $\varepsilon_\mathrm{T}$ and $\varepsilon_\mathrm{V}$ by a series expansion to the first order. We denote the approximate quantities calculated neglecting the light-time effects, that is for $\varepsilon_\mathrm{T}=\varepsilon_\mathrm{V}=0$, with a tilde. Substituting $f_\mathrm{T}$ from Eq. (\ref{eq:ftv-epsilon}) to the definition of the distance factor (\ref{eq:fd}), we get
\begin{equation}
  f_\mathrm{D}=\tilde{f}_\mathrm{D}-\tilde{f}_\mathrm{D}\tilde{\mu}_r\varepsilon_\mathrm{T}t\,.
\end{equation}
It follows from Eq. (\ref{eq:prop_u}) that the propagated barycentric position can be written as
\begin{equation}\label{eq:u-significance}
  \vec{u}=\vec{\tilde{u}}+\vec{\tilde{\mu}}\tilde{f}_\mathrm{D}^{-2}\varepsilon_\mathrm{T}t
\end{equation}
and formula (\ref{eq:prop_par}) gives the propagated parallax
\begin{equation}\label{eq:pi-significance}
  \varpi=\tilde{\varpi}-\tilde{\varpi}\tilde{\mu}_r\varepsilon_\mathrm{T}t\,.
\end{equation}
Expansion of the product $f_\mathrm{D}^3f_\mathrm{V}$, which appears in the formula of the propagated proper motion (\ref{eq:mu_prop}), to the first order in $\varepsilon_\mathrm{T}$ and $\varepsilon_\mathrm{V}$ gives $\tilde{f}_\mathrm{D}^3-3\tilde{f}_\mathrm{D}^3\tilde{\mu}_r\varepsilon_\mathrm{T}t+\tilde{f}_\mathrm{D}^3\varepsilon_\mathrm{V}$.
Since $\varepsilon_\mathrm{T}$ and $\varepsilon_\mathrm{V}$ are of the same order-of-magnitude, $\tilde{f}_\mathrm{D}\approx 1$ and $\tilde{\mu}_rt\ll 1$, we can omit the second term to get
\begin{equation}
  f_\mathrm{D}^3f_\mathrm{V}=\tilde{f}_\mathrm{D}^3+\tilde{f}_\mathrm{D}^3\varepsilon_\mathrm{V}
  \quad \mathrm{and} \quad
  f_\mathrm{D}^2f_\mathrm{V}=\tilde{f}_\mathrm{D}^2+\tilde{f}_\mathrm{D}^2\varepsilon_\mathrm{V}\,.
\end{equation}
The propagated proper motions then become
\begin{align}
    \label{eq:mu-significance}
    \vec{\mu}&=
    \vec{\tilde{\mu}}+\vec{\mu}_0{\tilde{f}_\mathrm{D}}^3\varepsilon_\mathrm{V}\,, \\
    \label{eq:mur-significance}
    \mu_r&=
    \tilde{\mu}_r+\mu_{r0}\tilde{f}_\mathrm{D}^2\varepsilon_\mathrm{V}\,.
\end{align}
Putting $\tilde{f}_\mathrm{D}=1$ in Eqs.~(\ref{eq:u-significance}), (\ref{eq:pi-significance}),
(\ref{eq:mu-significance}), and (\ref{eq:mur-significance}), we readily obtain the effects of the
light-time on the astrometric parameters
\begin{equation}\label{eq:effects}
 \begin{aligned}
    \Delta\theta&=\frac{\mu^3\tau_\mathrm{A}}{2\varpi}t^2 \,,
    &\Delta\varpi&=\frac{\mu_r\mu^2\tau_\mathrm{A}}{2}t^2\,,\\
    \Delta\mu&=\frac{\mu^3\tau_\mathrm{A}}{\varpi}t\,,
    &\Delta\mu_r&=\frac{\mu_r\mu^2\tau_\mathrm{A}}{\varpi}t\,.
 \end{aligned}
\end{equation}
We note the following relations between the effects
\begin{equation}\label{eq:effects-relations}
    \Delta\theta=\frac{1}{2}t\Delta\mu\quad\mbox{and}\quad
    \Delta\varpi=\frac{1}{2}t\Delta\mu_r \, .
\end{equation}
It is instructive to express the effects in
terms of the physical parameters, including the effects in velocity:
\begin{equation}\label{eq:effects_physical}
 \begin{aligned}
    \Delta\theta&=\frac{1}{2}\frac{v_t}{c}\left(\frac{v_t}{b}\right)^2t^2 \,,
    &\Delta\varpi&=\frac{1}{2}\frac{A}{b}\frac{v_r}{c}\left(\frac{v_t}{b}\right)^2t^2\,,\\
    \Delta\mu&=\frac{v_t}{c}\left(\frac{v_t}{b}\right)^2t\,,
    &\Delta\mu_r&=\frac{v_r}{c}\left(\frac{v_t}{b}\right)^2t\,,\\
    \Delta v&=\frac{vv_t^2}{cb}t\,,
    &\Delta v_t&=\frac{v_t^3}{cb}t\,, \quad \Delta v_r=\frac{v_rv_t^2}{cb}t\,.
 \end{aligned}
\end{equation}
These relations lead to important conclusions about the behaviour of the effects. The effects on the
position and parallax are quadratic functions of time, while the effects on the proper motion and velocity
increases linearly with time. This confirms the conclusion drawn from the numerical calculations shown
in Fig.~\ref{fig:prop}. All the effects are roughly proportional to the third power of the space velocity,
while the dependence on distance is different for the velocities ($b^{-1}$),
position and proper motions ($b^{-2}$), and parallax ($b^{-3}$).

For the practical estimation of the effects we give the following formulae for the position,
\begin{equation*}
    \Delta\theta=\left(0.36~\mu\mathrm{as}\right)
    \times\left(\frac{v_t}{10^3~\mathrm{km~s}^{-1}}\right)^3
    \times\left(\frac{b}{1~\mathrm{kpc}}\right)^{-2}
    \times\left(\frac{t}{100~\mathrm{yr}}\right)^2\,,
\end{equation*}
and velocity,
\begin{multline*}
    \Delta v=\left(0.34\ \mathrm{m}\,\mathrm{s}^{-1}\right)
    \times\left(\frac{v}{10^3\ \mathrm{km}\,\mathrm{s}^{-1}}\right)
    \times\left(\frac{v_t}{10^3\ \mathrm{km}\,\mathrm{s}^{-1}}\right)^2\\
    \times\left(\frac{b}{1\ \mathrm{kpc}}\right)^{-1}
    \times\left(\frac{t}{100\ \mathrm{yr}}\right)\,.
\end{multline*}

\section{Applicability of the uniform rectilinear model}
\label{ss:model_applicability}

In this appendix we briefly consider the conditions under which stellar motion may be regarded as uniform.
A uniform motion implies absence of acceleration. In practice, however, accelerated motion may be treated
as uniform if observable effects of the acceleration are negligible compared to the required astrometric
accuracy. The effect of a constant acceleration $\vec{a}$ on the barycentric position of a star during a timespan $t$ is $\Delta\vec{b}\simeq \vec{a}t^2/2$. The corresponding change in the angular position
$\theta$ of the star is $\Delta\theta\simeq a_\perp t^2/(2b)$, where $a_\perp$ is
the tangential component of the acceleration. The motion may be regarded as uniform if
$|\Delta\theta|\ll\sigma_\theta$, the required astrometric accuracy in angular position after time $t$.
For the proper motion, we similarly have the condition $|\Delta\mu|\ll\sigma_\mu$, where
$\Delta\mu\simeq a_\perp t/b$. The former (positional) criterion is usually stricter since
$t$ is typically much greater than $2\sigma_\theta/\sigma_\mu$.

The acceleration along the line of sight, $a_\parallel$ (taken to be positive when directed away from the
SSB), causes a change in parallax by $\Delta\varpi\simeq -A a_\parallel  t^2/(2b^2)$, where $A$ is the
astronomical unit, and in radial velocity by $\Delta v_r\simeq a_\parallel t$. If $a_\perp$ and $a_\parallel$
are of similar magnitudes, we find that $|\Delta\varpi|$ is smaller than $|\Delta\theta|$ by a factor
$A/b\ll 1$, so the effect in parallax is never a limitation. On the other hand, under fairly realistic
assumptions it may happen that the acceleration effect is more important in radial velocity than in
position.

We do not consider here the acceleration caused by stellar or planetary companions, which affects
specific objects in a very specific way and may be very important. Indeed, as emphasized in the
introduction, one of the objectives of the uniform rectilinear hypothesis is precisely to enable the
detection of such cases. Rather, we need to consider accelerations that affect all, or most of,
the stars and which could therefore potentially render the model invalid as a general basis for
high-precision astrometric analyses. The most important such acceleration is caused by the
large-scale gravitational field of the Galaxy, i.e. the curvature of Galactic stellar orbits.

At the arbitrary point $\vec{b}$ in the Galaxy (relative to the SSB) the acceleration vector can be
estimated as $\vec{a}=-\nabla\psi$, where $\psi$ is some suitable model of the Galactic
potential \citep{BT2}. It should be recalled that the uniform rectilinear model refers to the
motion of stars relative to the SSB, and that the SSB itself is subject to some acceleration
$\vec{a}(\vec{0})$. The observable effects must therefore be evaluated for the differential
acceleration $\Delta\vec{a}=\vec{a}(\vec{b})-\vec{a}(\vec{0})$, and the quantities $a_\parallel$ and
$a_\perp$ discussed above are therefore the components of $\Delta\vec{a}$ along and
perpendicular to the line of sight.%
\footnote{The acceleration of the SSB causes some observable effects on the proper motions
of all objects due to the slowly changing secular aberration
\citep{bastian1995, kovalevsky2003, liu+2013}. Studies of Galactic motions should in principle
be made in a galactocentric reference system, and the transformation from barycentric
quantities needs to take this effect into account as well as the secular aberration itself
(for the positions). This is not further discussed here.}
In a smooth potential both components vanish as $b\rightarrow 0$.

Rather than using a (rather uncertain) global potential model, however, it is more illuminating
to analyse the differential effects based on a few relatively well-determined structural Galactic
parameters. We assume an axisymmetric potential in galactocentric cylindrical coordinates $(R,z)$
and consider separately the acceleration components in the Galactic plane (along $R$) and
perpendicular to it (along $z$). To avoid confusion with the $b$ denoting a star's distance
from the SSB, we subsequently use $\glat$ to denote Galactic latitude, and $\glon$ for the longitude.

\paragraph{Acceleration in the Galactic plane}
In the axisymmetric approximation the acceleration in the Galactic plane
is directed towards the Galactic centre and of magnitude $a=V(R)/R^2$, where $V(R)$
denotes the circular velocity as radial distance $R$. The Sun
is currently located close to the Galactic plane at a radius $R_0\simeq 8.4$~kpc from the
Galactic centre, where the circular velocity is $V_0\equiv V(R_0)\simeq 254$~km~s$^{-1}$
\citep{reid+2009}. The expected acceleration at the location of the Sun is therefore
$a_0=V_0^2/R_0\simeq 2.5\times 10^{-10}$~m~s$^{-2}$.

The effects in position and proper motion are proportional to $a_\perp/b$, which in a smooth
potential become distance-independent for sufficiently small $b$, that is in the solar neighbourhood.
It is interesting to derive the corresponding local approximations for the acceleration components.
This can be done in complete analogy with the well-known derivation of the Oort formulae for
the radial and tangential velocities of circular motions in terms of the Oort constants $A$ and $B$
\citep[e.g.][]{bm1998}. With $a(R)$ denoting the acceleration towards the Galactic centre
at radius $R$, we find
\begin{equation}\label{eq:aR}
a_\perp/b = -E\sin 2\glon \, , \quad
a_\parallel/b = F + E\cos 2\glon \, ,
\end{equation}
where $\glon$ is the Galactic longitude of the star, as seen from the Sun, and
\begin{equation}
E = \frac{1}{2}\left( \frac{a_0}{R_0} - \left.\frac{\text{d}a}{\text{d}R}\right|_{R=R_0}\right) \, , \quad
F = -\frac{1}{2}\left( \frac{a_0}{R_0} - \left.\frac{\text{d}a}{\text{d}R}\right|_{R=R_0}\right) \,
\end{equation}
are constants analogous to $A$ and $B$ in the Oort formulae. Using $a(R)=V(R)/R^2$ we can in fact
express $E$ and $F$ in terms of the Oort constants as
\begin{equation}
E =  2A(A-B)\, , \quad
F =  (A+B)(A-B) \, .
\end{equation}
In Eq.~(\ref{eq:aR}) we take $a_\perp$ to be positive in the direction of increasing $\glon$.
Since the Galactic rotation curve is nearly flat ($A+B=0$), we have $F\simeq 0$ and
$E\simeq\Omega_0^2$, where $\Omega_0=V_0/R_0\simeq 9.8\times 10^{-16}$~s$^{-1}$ is the
circular angular velocity at the Sun. In the solar neighbourhood, the effect of the curvature of
Galactic orbits on the position after time $t$ can therefore be estimated as
\begin{equation}\label{eq:delta-theta}
\Delta\theta \simeq -\frac{1}{2} \left(\Omega_0 t\right)^2 \sin 2\glon \simeq
(-1.0~\mu\text{as})\times\left(\frac{t}{100~\text{yr}}\right)^2\times\sin 2\glon \, ,
\end{equation}
thus negligible at the 1~$\mu$as precision for time intervals up to 100~yr.
The corresponding effect on the radial velocity is
\begin{multline}
\Delta v_r \simeq \Omega_0^2 t b \cos 2\glon \\
\simeq (0.09~\text{m~s}^{-1})\times\left(\frac{t}{100~\text{yr}}\right)\times
\left(\frac{b}{1~\text{kpc}}\right)\times\cos 2\glon \, ,
\end{multline}
where we have again assumed a flat rotation curve.

Beyond the solar neighbourhood, e.g. at distances of the order of $R_0$ from the Sun, the
differential acceleration is of the order of the solar acceleration, or $a_0=\Omega_0^2R_0$.
The astrometric effects, being proportional to $a_\perp/b\sim a_0/R_0=\Omega_0^2$, are
therefore of the same order of magnitude as computed above for the solar neighbourhood.

\paragraph{Acceleration perpendicular to the Galactic plane}
In the solar neighbourhood, the component of the acceleration perpendicular to the
Galactic plane, at distance $z$ above the plane, is approximately given by
$a(z) = -2\pi G\Sigma(z)$, where $\Sigma(z)=\int_{-z}^z\rho(z')\,\text{d}z'$
is the surface density within $\pm z$ of the Galactic plane. Within a few hundred pc
from the Sun we can assume an approximately constant mass density $\rho_0$,
yielding $a(z)\simeq -Kz$ where $K=4\pi G\rho_0$ is the square of the angular
frequency of the oscillations in $z$. The acceleration relative to the
SSB follows the same formula if $z$ is interpreted as the vertical coordinate of
the star relative to the Sun, that is, $z=b\sin\glat$. For the components of $a(z)$
perpendicular to and along the line of sight we readily find
\begin{equation}\label{eq:az}
a_\perp/b = -{\textstyle\frac{1}{2}}K \sin 2\glat \, , \quad
a_\parallel/b = -{\textstyle\frac{1}{2}}K + {\textstyle\frac{1}{2}}K\cos 2\glat \, ,
\end{equation}
where $a_\perp$ is positive in the direction of increasing $\glat$.
Using $\rho_0\simeq 0.1~M_\odot~\text{pc}^{-3}$ \citep{holmberg+flynn2000},
we have $K\simeq 5.7\times 10^{-30}$~s$^{-2}$, and the accumulated
effects in position and radial velocity after time $t$ can then be estimated as
\begin{equation}
\Delta\theta \simeq -\frac{1}{4} K t^2 \sin 2\glat \simeq
(-2.9~\mu\text{as})\times\left(\frac{t}{100~\text{yr}}\right)^2\times\sin 2\glat \, ,
\end{equation}
and
\begin{multline}
\Delta v_r \simeq -K t b \sin^2\glat \\
\simeq (-0.28~\text{m~s}^{-1})\times\left(\frac{t}{100~\text{yr}}\right)\times
\left(\frac{b}{1~\text{kpc}}\right)\times \sin^2\glat \, .
\end{multline}
These approximations are valid for distances $b$ up to a few hundred pc,
beyond which the effects may be considerably smaller.

The effects of the acceleration perpendicular to the Galactic plane are therefore
more important than the radial acceleration, which simply reflects the shorter
oscillation period in the $z$ direction, $2\pi K^{-1/2}\simeq 84$~Myr,
compared to the circular period $2\pi/\Omega_0\simeq 200$~Myr.
However, the general conclusion is that Galactic accelerations are negligible at
micro-arcsecond accuracy over time periods of at least 50~yr.

\end{document}